\documentclass[sigconf]{acmart}

\AtBeginDocument{%
  \providecommand\BibTeX{{%
    \normalfont B\kern-0.5em{\scshape i\kern-0.25em b}\kern-0.8em\TeX}}}

\setcopyright{rightsretained}
\copyrightyear{2021}
\acmYear{2021}

\acmSubmissionID{asiafp199}

\usepackage{xspace}
\usepackage{booktabs}
\usepackage{enumerate}
\usepackage{verbatim}
\usepackage{amsmath}
\usepackage{amsfonts}
\usepackage{graphicx}
\usepackage{subcaption}
\usepackage{slashbox}
\usepackage[linesnumbered,ruled]{algorithm2e}
\usepackage{siunitx}
\usepackage{float}
\usepackage{listings}
\usepackage{fancyvrb}
\usepackage[export]{adjustbox}

\begin{document}

\fancyhead{}

\newif\ifdraft
\draftfalse

\newif\ifpublishable
\publishabletrue

\ifdraft
  \newcommand{\todocolor}[1]{\textcolor{red}{#1}}
\else
  \newcommand{\todocolor}[1]{}
\fi
\newcommand{\mahmood}[1]{\todocolor{[[Mahmood: #1]]}}
\newcommand{\keane}[1]{\todocolor{[[Keane: #1]]}}
\newcommand{\lujo}[1]{\todocolor{[[Lujo: #1]]}}
\newcommand{\mike}[1]{\todocolor{[[Mike: #1]]}}
\newcommand{\saurabh}[1]{\todocolor{[[Saurabh: #1]]}}
\newcommand\note[1]{\todocolor{[[Note: #1]]}}
\newcommand\todo[1]{\todocolor{[[#1]]}}

\newcommand{\algref}[1]{\mbox{Alg.~\ref{#1}}}
\newcommand{\secref}[1]{\mbox{Sec.~\ref{#1}}\xspace}
\newcommand{\secsref}[2]{\mbox{Sec.~\ref{#1}--\ref{#2}}\xspace}
\newcommand{\figref}[1]{\mbox{Fig.~\ref{#1}}}
\newcommand{\figrefs}[2]{\mbox{Figs.~\ref{#1}--\ref{#2}}\xspace}
\newcommand{\tabref}[1]{\mbox{Table~\ref{#1}}}
\newcommand{\listref}[1]{\mbox{Listing~\ref{#1}}}
\newcommand{\appref}[1]{\mbox{App.~\ref{#1}}}
\newcommand{\eqnref}[1]{Eqn.~\ref{#1}\xspace}
\newcommand{\eqnsref}[2]{Eqns.~\ref{#1}--\ref{#2}\xspace}

\newfloat{lstfloat}{htbp}{lop}
\floatname{lstfloat}{Listing}
\def\lstfloatautorefname{Listing} 
\lstdefinestyle{customstyle}{
  breaklines=true,
  frame=single,
  basicstyle=\small,
  numbers=left,
  numberstyle=\tiny,
  columns=fullflexible,
  mathescape=true,
  escapechar={^},
  linewidth=0.85\columnwidth,
  xleftmargin=0.09\columnwidth
}

\makeatletter
\DeclareRobustCommand{\varname}[1]{\begingroup\newmcodes@\mathit{#1}\endgroup}
\makeatother

\renewcommand{\ml}[0]{ML}
\newcommand{\dnn}[0]{DNN}
\newcommand{\lpnorm}[1]{$L_{#1}$}
\newcommand{\cwloss}[0]{\textit{Loss}_\textit{cw}}
\newcommand{\dnnloss}[0]{\mathbb{L}_{\dnnFuncBE}}
\newcommand{\dnnFuncBE}[0]{\mathbb{F}} 
\newcommand{\dnnFuncAE}[0]{\mathbb{H}}  
\newcommand{\dnnLogits}[0]{\mathbb{L}}
\newcommand{\embedFunc}[0]{\mathbb{E}}
\newcommand{\layerFunc}[0]{\ell}
\newcommand{\normFunc}[0]{\mathit{norm}}
\newcommand{\IPR}[0]{$\mathit{IPR}$}
\newcommand{\Disp}[0]{$\mathit{Disp}$}
\newcommand{\Kreuk}[0]{$\mathit{Kreuk}$}
\newcommand{\potency}[0]{potency}
\newcommand{\coverage}[0]{coverage}
\newcommand{\equivnt}[0]{\mathit{Eqv}}
\newcommand{\swap}[0]{\mathit{Regs}}
\newcommand{\reorder}[0]{\mathit{Ord1}}
\newcommand{\preserv}[0]{\mathit{Ord2}}
\newcommand{\pe}[0]{$\mathit{PE}$}
\newcommand{\malconv}[0]{$\mathit{MalConv}$}
\newcommand{\avastnet}[0]{$\mathit{AvastNet}$}
\newcommand{\ember}[0]{$\mathit{Endgame}$}
\newcommand{\vtfeed}[0]{$\mathit{VTFeed}$}
\newcommand{\kaggle}[0]{$\mathit{Kaggle}$}

\newcommand{\parheading}[1]{\textbf{\textit{#1}}~\hspace{2pt}}

\renewcommand{\textasciigrave}{'\xspace}

\date{}

\title{Malware Makeover: Breaking ML-based Static Analysis by Modifying Executable Bytes}

\acmConference[ASIA CCS '21] {Proceedings of the 2021 ACM Asia Conference on Computer and Communications Security}{June 7--11, 2021}{Hong Kong, Hong Kong}
\acmBooktitle{Proceedings of the 2021 ACM Asia Conference on Computer and Communications Security (ASIA CCS '21), June 7--11, 2021, Hong Kong, Hong Kong}
\acmPrice{}
\acmISBN{978-1-4503-8287-8/21/06}
\acmDOI{10.1145/3433210.3453086}

\ifpublishable
\author{Keane Lucas}
\email{keanelucas@cmu.edu}
\affiliation{
  \institution{Carnegie Mellon University}
}
\author{Mahmood Sharif}
\email{mahmoods@vmware.com}
\affiliation{
  \institution{Tel Aviv University and VMware}
}
\author{Lujo Bauer}
\email{lbauer@cmu.edu}
\affiliation{
  \institution{Carnegie Mellon University}
}
\author{Michael K.\ Reiter}
\email{michael.reiter@duke.edu}
\affiliation{
  \institution{Duke University}
}
\author{Saurabh Shintre}
\email{saurabh.shintre@nortonlifelock.com}
\affiliation{
  \institution{NortonLifeLock Research Group}
}
\else
\author{anon}
\fi


\begin{abstract}
  
Motivated by the transformative impact of deep neural
networks (\dnn{}s) in various domains, researchers and
anti-virus vendors have proposed \dnn{}s for malware
detection from raw bytes that do not require manual feature
engineering. In this work, we propose an attack that
interweaves binary-diversification techniques and
optimization frameworks to mislead such \dnn{}s while
preserving the functionality of binaries. Unlike prior
attacks, ours manipulates instructions that are a functional
part of the binary, which makes it particularly challenging
to defend against. We evaluated our attack against three
\dnn{}s in white- and black-box settings, and found that it
often achieved success rates near 100\%. Moreover, we found
that our attack can fool some commercial anti-viruses, in
certain cases with a success rate of 85\%. We explored
several defenses, both new and old, and identified some that
can foil over 80\% of our evasion attempts. However, these
defenses may still be susceptible to evasion by attacks, and
so we advocate for augmenting malware-detection systems with
methods that do not rely on machine learning.

\end{abstract}




\maketitle

\section{Introduction}

Modern malware detectors, both academic
(e.g.,~\cite{Arp14Drebin, Jindal19Neurlux}) and commercial
(e.g.,~\cite{Cylance, Symantec}), increasingly rely on
machine learning (\ml{}) to classify executables as benign
or malicious based on features such as imported libraries
and API calls. In the space of static malware detection,
where an executable is classified prior to its execution,
recent efforts have
proposed deep neural networks (\dnn{}s) that detect malware
from binaries' raw byte-level representation, with
effectiveness similar to that of detectors based on
hand-crafted features selected through tedious manual
processing~\cite{Krvcal18AvastNet, Raff18MalConv}.

As old techniques for obfuscating and packing malware (see
\secref{sec:results}) are rendered ineffective in the face
of static \ml{}-based detection, recent advances in adversarial
\ml{} might provide a new opening for attackers to bypass
detectors. Specifically, \ml{} algorithms, including
\dnn{}s, have been shown vulnerable to adversarial
examples---modified inputs that resemble normal inputs but
are intentionally designed to be misclassified. For
instance, adversarial examples can enable attackers to
impersonate users that are enrolled in face-recognition
systems~\cite{Sharif16AdvML, Sharif19AGNs}, fool street-sign
recognition algorithms into misclassifying street
signs~\cite{Evtimov17Signs}, and trick voice-controlled
interfaces to misinterpret commands~\cite{Cisse17Houdini,
Qin19Speech, Schon18AdvSound}.

In the malware-detection domain, the attackers' goal is to
alter programs to mislead \ml{}-based malware detectors to
misclassify malicious programs as benign or vice versa.
In doing so, attackers face a non-trivial constraint: in
addition to misleading the malware detectors, 
alterations to a program must not change its functionality. For
example, a keylogger altered to evade being detected as
malware should still carry out its intended function,
including invoking necessary APIs, accessing sensitive
files, and exfiltrating information. This constraint is
arguably more challenging than ones imposed by other domains
(e.g., evading image recognition without making changes
conspicuous to humans~\cite{Evtimov17Signs, Sharif16AdvML,
Sharif19AGNs}) as it is less amenable to being encoded into
traditional frameworks for generating adversarial examples,
and most changes to a program's raw binary are likely to
break a program's syntax or semantics. Prior work proposed
attacks to generate adversarial examples to fool static
malware detection \dnn{}s~\cite{Demetrio19FoolMD,
Kolos18Malware, Kreuk18Malware,
Suciu18Malware,Park19Malware} by
adding
adversarially crafted byte values in program regions that do not
affect execution (e.g., at the end of programs or between
sections). These attacks can be defended against by
eliminating the added content before classification
(e.g.,~\cite{Kruegel04Disas}); we confirm this empirically.

In contrast, we develop a new way to modify binaries to both
retain their functionality and mislead state-of-the-art
\dnn{}-based static malware
detectors~\cite{Krvcal18AvastNet, Raff18MalConv}. We
leverage binary-diversification tools---originally
proposed to defend against code-reuse attacks by
transforming program binaries to create diverse
variants~\cite{Koo16Rand, Pappas12Rand}---to evade
malware-detection \dnn{}s. While these tools preserve the
functionality of programs by design
(e.g., functionality-pre\-serv\-ing randomization), their
na\"ive application is insufficient to evade malware
detection. We propose optimization algorithms to guide the
transformations of binaries to fool malware-detection
\dnn{}s, both in settings where attackers have access to the
\dnn{}s' parameters (i.e., white-box) and ones where they
have no access (i.e., black-box). The algorithms we propose
can produce program variants that often fool \dnn{}s in
100\% of evasion attempts and, surprisingly, even evade some
commercial malware detectors (likely over-reliant on
\ml{}-based static detection), in some cases with success
rates as high as 85\%. Because our attacks transform
functional parts of programs, they are particularly
difficult to defend against, especially when augmented with
complementary methods to further deter static or dynamic analysis (as our methods alone
should have no effect on dynamic analysis). We explore
potential mitigations to our attacks (e.g., by normalizing
programs before classification~\cite{Armoun12Norm,
Christodorescu05Norm, Walenstein06Norm}), but identify their
limitation in thwarting adaptive attackers.

In a nutshell, the contributions of our paper are as
follows:
\begin{itemize}
\item We repair and extend prior binary-diversification
  implementations to iteratively yield
  candidate transformations. We
  also reconstruct them to be composable, more capable, and
  resource-efficient. The code is available
  online.\footnote{\url{https://github.com/pwwl/enhanced-binary-diversification}}
\item We propose a novel
  functionality-pre\-ser\-ving attack on \dnn{}s for static
  malware detection from raw bytes
  (\secref{sec:tech_approach}). The attack precisely
  composes the updated binary-diversification
  techniques, evades defenses against prior
  attacks, and applies to both white- and black-box
  settings.
\item We evaluate and demonstrate the effectiveness of the
  proposed attack in different settings, including against
  commercial malware detectors (\secref{sec:results}). We
  show that our attack effectively undermines ML-based
  static analysis, a significant component of
  state-of-the-art malware detection, while being robust to
  defenses that can thwart prior attacks. 
\item We explore the effectiveness of prior and new defenses
  against our proposed attack (\secref{sec:discussion}).
  While some defenses seem promising against specific
  variants of the attack, none explored neutralize our most
  effective attack, and they are likely vulnerable
  to adaptive attackers.
\end{itemize}

\section{Background and Related Work}
\label{sec:relwork}

We first discuss previous work on \dnn{}s that detect malware
by examining program binaries.
We then discuss research on
attacking and defending \ml{} algorithms generally, and
malware detection specifically. Finally, we provide 
background on binary randomization methods, which serve as
building blocks for our attacks.

\subsection{\dnn{}s for Static Malware Detection}
\label{sec:relwork:maldet}

We study attacks targeting two \dnn{} architectures for
detecting malware from the raw bytes of Windows binaries
(i.e., executables in Portable Executable
format)~\cite{Krvcal18AvastNet, Raff18MalConv}. The main
appeal of these \dnn{}s is that they achieve
state-of-the-art performance using automatically learned
features, instead of manually crafted features that require
tedious human effort (e.g.,~\cite{Arp14Drebin,
Kolter06Malware, Incer18RobustMD}). Due to their desirable
properties, computer-security companies use \dnn{}s similar
to the ones we study (i.e., ones that operate on raw bytes
and use a convolution architectures) for malware
detection~\cite{FireEyeDNNs}.

The \dnn{}s proposed by prior work follow standard
convolutional architectures similar to the ones used for
image classification~\cite{Krvcal18AvastNet, Raff18MalConv}.
Yet, in contrast to image classifiers that classify
continuous inputs, malware-detection
\dnn{}s classify discrete inputs---byte values
of binaries. To this end, the \dnn{}s were designed with
initial embedding layers that map each byte in the input to
a vector in $\mathbb{R}^8$. After the embedding, standard
convolutional and non-linear operations are performed by
subsequent layers.

\subsection{Attacking and Defending \ml{} Algorithms}

\parheading{Attacks on Image Classification}
Adversarial examples---inputs that are minimally perturbed
to fool \ml{} algorithms---have emerged as challenge to
\ml{}. Most prior attacks (e.g.,~\cite{Baluja17ATNs,
Biggio13Evasion, Carlini17Robustness, Gdfllw14ExpAdv,
Papernot16Limitations, Szegedy13NNsProps}) focused on \dnn{}s for image
classification, and on finding adversarial perturbations
that have small \lpnorm{p}-norm ($p$ typically
$\in\{0,2,\infty\}$) that lead to misclassification when
added to input images. By limiting perturbations to small
\lpnorm{p}-norms, attacks aim to ensure that the
perturbations are imperceptible to humans. Attacks are often
formalized as optimization processes; e.g., Carlini and
Wagner~\cite{Carlini17Robustness} proposed the following
formulation for finding adversarial perturbations that
target a class $c_t$ and have small \lpnorm{2}-norms:
$$\arg\min_{r}\  \cwloss(x+r,c_t) + \kappa\cdot||r||_{2}$$
where $x$ is the original image, $r$ is the perturbation,
and $\kappa$ is a parameter to tune the \lpnorm{2}-norm of
the perturbation. $\cwloss$ is a function that, when
minimized, leads $x+r$ to be (mis)classified as $c_t$. It is
roughly defined as:
$$  \cwloss(x+r,c_t) = \max_{c\neq c_t}
\{\dnnLogits_{c}(x+r) \} - \dnnLogits_{c_t}(x+r)$$ where
$\dnnLogits_{c}$ is the output for class $c$ at the logits
of the \dnn{}---the output of the one-before-last layer. Our
attacks use $\cwloss$ to mislead the malware-detection
\dnn{}s.

\parheading{Attacks on Static Malware Detection}
Modern malware detection systems often leverage both dynamic
and static analyses to determine
maliciousness~\cite{Symantec,Cylance,Jindal19Neurlux,Avast,TrendMicro}.
While in most cases an attacker would hence need to adopt
countermeasures against both of these types of analyses, in
other situations, such as potential attacks on end-user
systems protected predominantly through static-analysis
based anti-virus detectors~\cite{ClamAV,Vipre}, defeating a
static malware detector could be sufficient for an attacker
to achieve their goals. Even when a combination of static
and dynamic analyses is used for detecting malware, fooling
static analysis is necessary for an attack to succeed. Here
we focus on attacks that target \ml{}-based static analyzers
for detecting malware.

Multiple attacks were proposed to evade \ml{}-based malware
classifiers while preserving the malware's functionality.
Some (e.g.,~\cite{Dang17Bbox, Srndic14PDFfool,
Wagner02Mimicry, Xu16PDFfool}) tweak malware to mimic benign
files (e.g., adding benign code-snippets to malicious PDF
files). Others (e.g.,~\cite{Anderson18Malware,
Demetrio19FoolMD, Grosse17Malware, Hu17AdvAPIs,
Kolos18Malware, Kreuk18Malware, Suciu18Malware,
Park19Malware}) tweak malware using gradient-based
optimizations or generative methods (e.g., to find which
APIs to import). Still others combine mimicry and
gradient-based optimizations~\cite{Rosenberg18Malware}.

Differently from some prior work
(e.g.,~\cite{Anderson18Malware, Rosenberg18Malware,
Wagner02Mimicry}) that studied attacks against dynamic
\ml{}-based malware detectors, we explore attacks that
target \dnn{}s for malware detection from raw bytes (i.e.,
static detection methods). Furthermore, the attacks we
explore do not introduce adversarially crafted bytes to
unreachable regions of the binaries~\cite{Kolos18Malware,
Kreuk18Malware, Suciu18Malware} (which may be possible to
detect and sanitize statically, see
\secref{sec:results:attackwb}), or by mangling bytes in the
header of binaries~\cite{Demetrio19FoolMD} (which can be
stripped before classification~\cite{Ronen18MSRKaggle}).
Instead, our attacks transform actual instructions of
binaries in a functionality-preserving manner to achieve
misclassification. 

More traditionally, attackers use various obfuscation
techniques to evade malware detection.
Packing~\cite{Biondi18PackDet, Roundy13Packing, Szor05Art,
Ugarte15Pack}---compressing or encrypting binaries' code and
data, and then uncompressing or decrypting them at run
time---is commonly used to hide malicious content from
static detection methods. As we explain later
(\secref{sec:tech:threat}) we mostly consider unpacked
binaries in this work, as is typical for static
analysis~\cite{Biondi18PackDet, Krvcal18AvastNet}. Attackers
also obfuscate binaries by substituting instructions or
altering their control-flow graphs~\cite{Christo03Testing,
Christo05Semantics, Junod15ollvm, Szor05Art}. We demonstrate
that such obfuscation methods do not fool malware-detection
\dnn{}s when applied na\"ively (see
\secref{sec:results:rand}). To address this, our attacks
guide the transformation of binaries via stochastic
optimization techniques to mislead malware detection.

Pierazzi et al.\ formalized the process of adversarial
example generation in the problem space and used their
formalization to produce malicious Android apps that evade
detection~\cite{Pierazzi20PSAttacks}. Our attack fits the
most challenging setting they describe, where mapping the
problem space to features space is non-invertible and
non-differentiable.

Most closely related to our work is the recent work on
misleading \ml{} algorithms for authorship
attribution~\cite{Meng18FoolAuthor, Quiring19Author}. Meng
et al.\ proposed an attack to mislead authorship attribution
at the binary level~\cite{Meng18FoolAuthor}. Unlike the
attacks we propose, Meng et al.\ leverage weaknesses in
feature extraction and modify debug information and
non-loadable sections to fool the \ml{} models. Furthermore,
their method leaves a conspicuous footprint that the binary
was modified (e.g., by introducing multiple data and code
sections to the binaries). While this is potentially
acceptable for evading author identification, it may raise
suspicion when evading malware detection. Quiring et al.\
recently proposed an attack to mislead authorship
attribution from source code~\cite{Quiring19Author}. In a
similar spirit to our work, their attack leverages an
optimization algorithm to guide code transformations that
change syntactic and lexical features of the code (e.g.,
switching between \texttt{printf} and \texttt{cout}) to
mislead \ml{} algorithms for authorship attribution.

\parheading{Defending \ml{} Algorithms} Researchers are
actively seeking ways to defend against adversarial
examples. One line of work, called adversarial training,
aims to train robust models largely by augmenting the
training data with correctly labeled adversarial
examples~\cite{Gdfllw14ExpAdv, Kantchelian16ICML,
Kannan18ALP, Kurakin16AdvTrain, Madry17AdvTraining,
Szegedy13NNsProps}. Another line of work proposes algorithms
to train certifiably (i.e., provably) robust defenses
against certain attacks~\cite{Cohen19RandSmooth,
Kolter17Defense, Lecuyer19PixelDP, Mirman18Defense,
Zhang19PrincipDef}, though these defenses are limited to
specific types of perturbations (e.g., ones with small
\lpnorm{2}- or \lpnorm{\infty}-norms). Moreover, they often
do not scale to large models that are trained on large
datasets. As we show in \secref{sec:discussion}, amongst
other limitations, these defenses would also be too
expensive to practically mitigate our attacks. Some defenses
suggest that certain input transformations (e.g.,
quantization) can ``undo'' adversarial perturbations before
classification~\cite{Guo18ReformDef, Liao18ReformDef,
Meng17Magnet, Samangouei18DefGAN, Srini18ReformDef,
Xie18Denoise, Xu18Squeeze}. In practice, however, it has
been shown that attackers can adapt to circumvent such
defenses~\cite{Athalye18BeatCVPR, Athalye18Attack}.
Additionally, the input transformations that have been
explored in the image-classification domain cannot be
applied in the context of malware detection. Prior work has
also shown that attackers~\cite{Carlini17Bypass} can
circumvent methods for detecting the presence of attacks
(e.g.,~\cite{Feinman17Detector, Grosse17Detector,
Meng17Magnet, Metzen17Detector}). We expect that such
attackers can circumvent attempts to detect our attacks too.

Prior work proposed \ml{}-based malware-classification
methods designed to be robust against
evasion~\cite{Demontis17RobustMD, Incer18RobustMD}. However,
these methods either have low
accuracy~\cite{Incer18RobustMD} or target linear
classifiers~\cite{Demontis17RobustMD}, which are unsuitable
for detecting malware from raw bytes.

Fleshman et al.\ proposed to harden malware-detection
\dnn{}s by constraining parameter weights in the last layer
to non-negative values~\cite{Fleshman18Def}. Their approach
aims to prevent attackers from introducing additional
features to malware to decrease its likelihood of being
classified correctly. While this rationale holds for
single-layer neural networks (i.e., linear classifiers),
\dnn{}s with multiple layers constitute complex functions
where feature addition at the input may correspond to
feature deletion in deep layers. As a result of the
misalignment between the threat model and the defense, we
found that \dnn{}s trained with this defense are as
vulnerable to prior attacks~\cite{Kreuk18Malware} as
undefended \dnn{}s.

\subsection{Binary Rewriting and Randomization}
\label{sec:relwork:binrand}

Software diversification is an approach to produce diverse
binary versions of programs, all with the same
functionality, to resist different kinds of attacks, such as
memory corruption, code injection, and code
reuse~\cite{Larsen14SoK}. Diversification can be performed
on source code, during compilation, or by rewriting and
randomizing  programs' binaries. In this work, we build on
binary-level diversification techniques, as they have wider
applicability (e.g., self-spreading malware can use them to
evade detection without source-code
access~\cite{Moser07Opaque}). Nevertheless, we expect that
this work can be extended to work with different
diversification methods.

Binary rewriting takes many forms
(e.g.,~\cite{Harris05Dyninst, Koo16Rand, Koo18Rand,
Massalin87Superoptimizer, Pappas12Rand, Schkufza13Rewriting,
Wang16Uroboros}). Certain methods aim to speed up code via
expensive search through the space of equivalent
programs~\cite{Massalin87Superoptimizer,
Schkufza13Rewriting}. Other methods significantly increase
binaries' sizes, or leave conspicuous signs that rewriting
took place~\cite{Harris05Dyninst, Wang16Uroboros}. We build
on binary-randomization tools that have little-to-no effect
on the size or run time of randomized binaries, thus helping
our attacks remain stealthy~\cite{Koo16Rand, Pappas12Rand}.
We present these tools and our extensions thereof in
\secref{sec:tech:attack}.

\section{Technical Approach}
\label{sec:tech_approach}

Here we present the technical approach of our attack. Before
delving into the details, we initially describe the threat
model.

\subsection{Threat Model}
\label{sec:tech:threat}

We assume that the attacker has white-box or black-box
access to \dnn{}s for malware detection that receive raw
bytes of program binaries as input. In the white-box
setting, the attacker has access to the \dnn{}s'
architectures and weights and can efficiently compute the
gradients of loss functions with respect to the \dnn{}s'
input via forward and backward passes. On the other hand,
the attacker in the black-box setting may only query the
model with a binary and receive the probability estimate
that the binary is malicious.

The \dnn{}s' weights are fixed and \emph{cannot} be
controlled by attackers (e.g., by poisoning the training
data). The attackers use binary rewriting to manipulate the
raw bytes of binaries and cause misclassification while
keeping functionality intact. Namely, attackers aim mislead
the \dnn{}s while ensuring that the I/O behavior of program
and the order of syscalls remain the same after rewriting.
In certain practical settings (e.g., when both dynamic and
static detection methods are used~\cite{Szor05Art}) evading
static detection techniques as the \dnn{}s we study may be
insufficient to evade the complete stack of detectors.
Nonetheless, evading the static detection techniques in such
settings is \emph{necessary} for evading detection overall.
In \secref{sec:vttransfer}, we show that our attacks can
evade commercial detectors, some of which may be using
multiple detection methods.

Attacks may seek to cause malware to be misclassified as
benign or benign binaries to be misclassified as malware.
The former may cause malware to circumvent defenses and be
executed on a victim's machine. The latter induces false
positives, which may lead users to turn off or ignore the
defenses~\cite{Herley09Extern}. Our methods are applicable
to transform binaries in either direction, but we focus on
transforming malicious binaries in this paper.

As is common for static malware
detection~\cite{Biondi18PackDet, Krvcal18AvastNet}, we
assume that the binaries are unpacked. While adversaries may
attempt to evade detection via packing, our attack can act
as an alternative or a complementary evasion technique
(e.g., once packing is undone). Such a technique is
particularly useful as packer-detection
(e.g.,~\cite{Biondi18PackDet}) and unpacking
(e.g.,~\cite{Cheng18Unpack}) techniques improve. In fact, we
found that packing with a popular packer increases the
likelihood of detection for malicious binaries (see
\secref{sec:vttransfer}), thus further motivating the need
for complementary evasion measures.

As is standard for \ml{}-based malware detection from raw
bytes in particular (\secref{sec:relwork:maldet}), and for
classification of inputs from discrete domains in general
(e.g.,~\cite{Le14Doc2Vec}), we assume that the first layer
of the \dnn{} is an embedding layer. This layer maps each
discrete token from the input space to a vector of real
numbers via a function $\embedFunc{}(\cdot)$. When computing
the \dnn{}'s output $\dnnFuncBE(x)$ on an input binary $x$,
one first computes the embeddings and feeds them to the
subsequent layers. Thus, if we denote the composition of the
layers following the embedding by $\dnnFuncAE(\cdot)$, then
$\dnnFuncBE(x)=\dnnFuncAE(\embedFunc(x))$. While the \dnn{}s
we attack contain embedding layers, our attacks conceptually
apply to \dnn{}s that do not contain such layers.
Specifically, for a \dnn{} function
$\dnnFuncBE(x)=\layerFunc_{n-1}(\dots
\layerFunc_{i+1}(\layerFunc_i(\dots
\layerFunc_0(x)\dots))\dots)$ for which the errors can be
propagated back to the $(i+1)^{\mathit{th}}$ layer, the
attack presented below can be executed by defining
$\embedFunc(x)=\layerFunc_i(\dots \layerFunc_0(x)\dots)$.

\subsection{Functionality-Preserving Attack}
\label{sec:tech:attack}

The attack we propose iteratively transforms a binary $x$ of
class $y$ ($y$=0 for benign binaries and $y$=1 for malware)
until misclassification occurs or a maximum number of
iterations is reached. To keep the binary's functionality
intact, only functionality preserving transformations are
used. In each iteration, the attack determines the subset of
transformations that can be safely used on each function in
the binary. The attack then randomly selects a
transformation from each function-specific subset and
enumerates candidate byte-level changes. Each candidate set
of changes is mapped to its corresponding gradient. The
changes are only applied if this gradient has positive
cosine similarity with the target model's loss gradient.

\begin{algorithm}
  \small
  \caption{White-box attack.}
  \label{alg:attack}

  \SetKwData{Left}{left}\SetKwData{This}{this}\SetKwData{Up}{up}
  \SetKwFunction{Union}{Union}\SetKwFunction{FindCompress}{FindCompress}
  \SetKwInOut{Input}{Input}\SetKwInOut{Output}{Output}

  \Input{$\dnnFuncBE=\dnnFuncAE(\embedFunc(\cdot))$, $\dnnloss$, $x$, $y$, $\mathit{niters}$}
  \Output{$\hat{x}$}
  \BlankLine
  $i\leftarrow 0$\;
  $\hat{x}\leftarrow \mathit{RandomizeAll}(x)$\;
  \While{$\dnnFuncBE(\hat{x})=y$ and  $i<\mathit{niters}$}{
    \For{$f \in \hat{x}$}{
      $\hat{e}\leftarrow \embedFunc(\hat{x})$\;
      $g\leftarrow \frac{\partial \dnnloss(\hat{x},y)}{\partial \hat{e}}$\;
      $o\leftarrow\mathit{RandomTransformationType}()$\;
      $\tilde{x}\leftarrow \mathit{RandomizeFunction}(\hat{x},f,o)$\;
      $\tilde{e}\leftarrow \embedFunc(\tilde{x})$\;
      $\delta_f=\tilde{e}_{f}-\hat{e}_{f}$\;
      \If{$g_f\boldsymbol{\cdot}\delta_f>0$}{
        $\hat{x}\leftarrow \tilde{x}$\;
      }
    }
    $i \leftarrow i+1$\;
  }
  \Return{$\hat{x}$}\;
\end{algorithm}

\algref{alg:attack} presents the pseudocode of the attack in
the white-box setting. The algorithm starts with a random
initialization. This is manifested by transforming all the
functions in the binary in an undirected way. Namely, for
each function in the binary, a transformation type is
selected at random from the set of available transformations
and applied to that function without consulting
loss-gradient similarity. When there are multiple ways to
apply the transformation to the function, one is chosen at
random. The algorithm then proceeds to further transform the
binary using our gradient-guided method for up to
$\mathit{niters}$ iterations. 

Each iteration starts by computing the embedding of the
binary to a vector space, $\hat{e}$, and the gradient, $g$,
of the \dnn{}'s loss function, $\dnnloss{}$, with respect to
the embedding. Particularly, we use the $\cwloss$, presented
in \secref{sec:relwork}, as loss function.  Because the true
value of $g$ is affected by any committed function change
and could be unreliable after transforming many preceding
functions in large files, it is recalculated prior to
transforming each function (lines 5--6).

Ideally, to move the binary closer to misclassification, we
would manipulate the binary so that the difference of its
embedding from $\hat{e}+\alpha{}g$ (for some scaling factor
$\alpha{}$) is minimized (see prior work for
examples~\cite{Kolos18Malware, Kreuk18Malware}). However, if
applied naively, such manipulation would likely cause the
binary to be ill-formed or change its functionality.
Instead, we transform the binary via
functionality-preserving transformations. As the
transformations are stochastic and may have many possible
outcomes (in some cases, more than can be feasibly
enumerated), we cannot precisely estimate their impact on
the binary a priori. Therefore, we implement the
transformation of each function, $f$, as the acceptance or
denial of candidate functionality-preserving transformations
we iteratively generate throughout the function, where we
apply a transformation only if it shifts the embedding in a
direction similar to $g$ (lines 5--13). More concretely, if
$g_f$ is the gradient with respect to the embedding of the
bytes corresponding to $f$, and $\delta_f$ is the difference
between the embedding of $f$'s bytes after the attempted
transformation and its bytes before, then each small
candidate transformation is applied only if the cosine
similarity (or, equivalently, the dot product) between $g_f$
and  $\delta_f$ is positive. Other optimization methods
(e.g., genetic programming~\cite{Xu16PDFfool}) and
similarity measures (e.g., similarity in the Euclidean
space) that we tested did not perform as well. 

If the input was continuous, it would be possible to
perform the same attack in a black-box setting after
estimating the gradients by querying the model
(e.g.,~\cite{Ilyas18Bbox}). In our case, however, it is not
possible to estimate the gradients of the loss with respect
to the input, as the input is discrete. Therefore, the
black-box attack we propose follows a general hill-climbing
approach (e.g.,~\cite{Srndic14PDFfool}) rather than gradient
ascent. The black-box attack is conceptually similar to the
white-box one, and differs only in the method of checking
whether to apply attempted transformations: Whereas the
white-box attack uses gradient-related information to decide
whether to apply a transformation, the black-box attack
queries the model after attempting to transform a function
and accepts the transformation only if the probability of
the target class increases.

\begin{figure*}[!ht]
  \centering

  \begin{subfigure}{0.195\textwidth}
    \includegraphics[width=\textwidth]{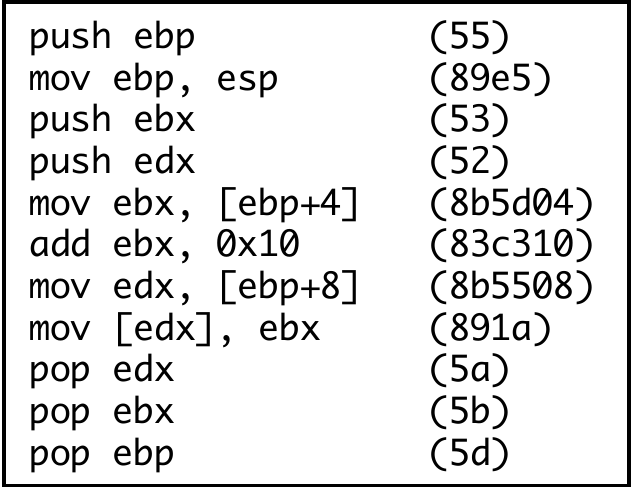}
    \caption{Original}
  \end{subfigure}
  \begin{subfigure}{0.195\textwidth}
    \centering
    \includegraphics[width=\textwidth]{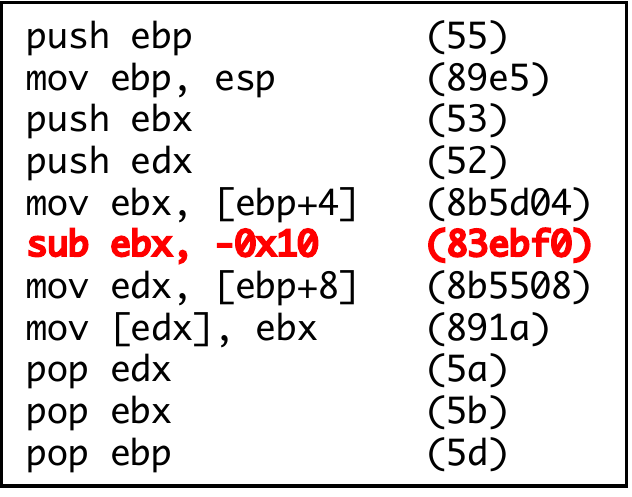}
    \caption{Equivalent instructions}
  \end{subfigure}
  \begin{subfigure}{0.195\textwidth}
    \centering
    \includegraphics[width=\textwidth]{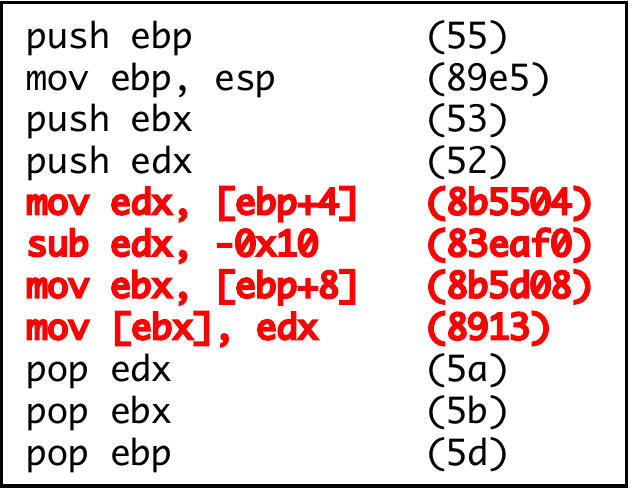}
    \caption{Register reassignment}
  \end{subfigure}
  \begin{subfigure}{0.195\textwidth}
    \centering
    \includegraphics[width=\textwidth]{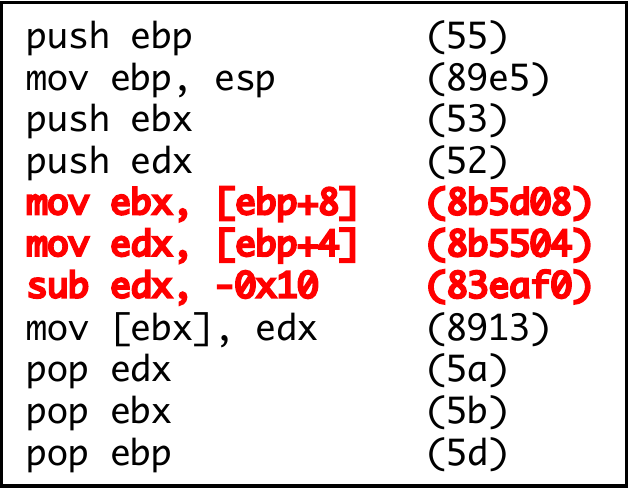}
    \caption{Instruction reordering}
  \end{subfigure}
  \begin{subfigure}{0.195\textwidth}
    \centering
    \includegraphics[width=\textwidth]{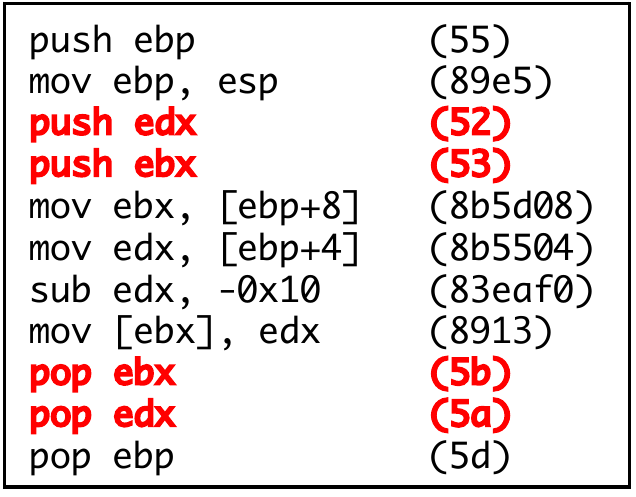}
    \caption{Register preservation}
  \end{subfigure}
  
  \caption{\label{fig:ipr}An illustration of \IPR{}. We show how the original
    code (a) changes after replacing instructions with equivalent ones (b),
    reassigning registers (c), reordering instructions (d), and changing the
    order of instructions that preserve register values (e). We provide the hex
    encoding of each instruction to its right. The affected instructions are
    boldfaced and colored in red.}
\end{figure*}

\parheading{Transformation Types}
We consider two families of transformation
types~\cite{Koo16Rand, Pappas12Rand}. As the first family,
we adopt and extend transformation types proposed for
in-place randomization (\IPR{})~\cite{Pappas12Rand}. Given a
binary to randomize, Pappas et al.\ proposed to disassemble
it and identify functions and basic blocks, statically
perform four types of transformations that preserve
functionality, and then update the binary accordingly from
the modified assembly. The four transformation types are:
\emph{1)} replacing instructions with equivalent ones of the
same length (e.g., \texttt{sub eax,4} $\rightarrow$
\texttt{add eax,-4});
\emph{2)} 
reassigning registers within functions or sets of basic
blocks (e.g., swapping all instances of \texttt{ebx} and
\texttt{ecx}) if this does not affect code that follows;
\emph{3)}
reordering instructions using a dependence graph to ensure
that no instruction appears before one it depends on; and
\emph{4)} altering the order in which register values are
pushed to and popped from the stack to preserve them across
function calls. To maintain the semantics of the code, the
disassembly and transformations are performed conservatively
(e.g., speculative disassembly, which is likely to
misidentify code, is avoided). \IPR{} does not alter
binaries' sizes and has no measurable effect on their run
time~\cite{Pappas12Rand}. \figref{fig:ipr} shows examples of
transforming code via \IPR{}.

The original implementation of Pappas et al.\ was unable to
produce the majority of functionally equivalent binary
variants that should be achievable under the four
transformation types. Thus, we extended and improved the
implementation in various ways. First, we enabled the
transformations to compose: unlike Pappas et al.'s
implementation, our implementation allows us to iteratively
apply different transformation types to the same function.
Second, we apply transformations more conservatively to
ensure that the functionality of the binaries is preserved
(e.g., by not replacing \texttt{add} and \texttt{sub}
instructions if they are followed by instructions that read
the flags register). Third, compared to the previous
implementation, ours handles a larger number of instructions
and function-calling conventions. In particular, our
implementation can rewrite binaries containing additional
instructions (e.g., \texttt{shrd}, \texttt{shld},
\texttt{ccmove}) and less common calling conventions (e.g.,
nonstandard returns via increment of \texttt{esp} followed
by a \texttt{jmp} instruction). Last, we fixed significant
bugs in the original implementation. These bugs include
incorrect checks for writes to memory after reads, as well
as memory leaks which required routine experiment restarts.

The second family of transformation types that we build on
is based on code displacement (\Disp{})~\cite{Koo16Rand}.
Similarly to \IPR{}, \Disp{} begins by conservatively
disassembling the binary. The original idea of \Disp{} is to
break potential gadgets that can be leveraged by code-reuse
attacks by moving code to a new executable section. The
original code to be displaced has to be at least five bytes
in size so that it can be replaced with a \texttt{jmp}
instruction that passes control to the displaced code. If
the displaced code contains more than five bytes, the bytes
after the \texttt{jmp} are replaced with trap instructions
that terminate the program; these would be executed if a
code-reuse attack is attempted. In addition, another
\texttt{jmp} instruction is appended to the displaced code
to pass control back to the instruction that should follow.
Any displaced instruction that uses an address relative to
the instruction pointer (i.e., \texttt{IP}) register is also
updated to reflect the new address after displacement.
\Disp{} has a minor effect on binaries' sizes ($\sim$2\%
increase on average) and causes a small amount of run-time
overhead ($<$1\% on average)~\cite{Koo16Rand}.

\begin{figure}
\centering
\begin{lstlisting}[style=customstyle]
$\mathit{S}$ $\rightarrow$  $\mathit{Atom}$ | $\mathit{S}\cdot \mathit{S}$ |
      ^\texttt{bswp r}^ $\cdot$ $\mathit{S}$ $\cdot$ ^\texttt{bswp r}^ |
      ^\texttt{xchg rh, rl}^ $\cdot$ $\mathit{S}$ $\cdot$ ^\texttt{xchg rh, rl}^ |
      ^\texttt{push r}^ $\cdot$ $\mathit{S_{r}}$ $\cdot$ ^\texttt{pop r}^ |
      ^\texttt{pushfd}^ $\cdot$ $\mathit{S_{ef}}$ $\cdot$ ^\texttt{popfd}^ 
$\mathit{Atom}$ $\rightarrow$ $\Phi$ | ^\texttt{nop}^ | ^\texttt{mov r, r}^
$\mathit{S_{r}}$ $\rightarrow$  $\mathit{S}$ | $\mathit{S_{r}}$ $\cdot$ $\mathit{S_{r}}$ | ^\texttt{pushfd}^ $\cdot$ $\mathit{S_{ef,r}}$ $\cdot$ ^\texttt{popfd}^
$\mathit{S_{ef}}$ $\rightarrow$  $\mathit{S}$ | $\mathit{S_{ef}}$ $\cdot$ $\mathit{S_{ef}}$ |
        ^\texttt{arth r, v}^ $\cdot$ $\mathit{S_{ef}}$ $\cdot$  ^\texttt{invarth r, v}^ |  
        ^\texttt{push r}^ $\cdot$ $\mathit{S_{ef,r}}$ $\cdot$ ^\texttt{pop r}^  
$\mathit{S_{ef,r}}$ $\rightarrow$  $\mathit{S}$ | $\mathit{S_r}$ | $\mathit{S_{ef}}$ | $\mathit{S_{ef,r}}$ $\cdot$ $\mathit{S_{ef,r}}$  | 
        ^\texttt{arth r, v}^ $\cdot$ $\mathit{S_{ef,r}}$ |
        ^\texttt{logic r, v}^ $\cdot$ $\mathit{S_{ef,r}}$ 
\end{lstlisting}
\caption{\label{fig:semnops}A context-free grammar for generating semantic
  nops. $\mathit{S}$ is the starting symbol; $\Phi$ the empty string;
  \texttt{arth} indicates an arithmetic operation (specifically, \texttt{add},
  \texttt{sub}, \texttt{adc}, or \texttt{sbb}); \texttt{invarth} indicates its
  inverse; \texttt{logic} indicates a logical operation (specifically, \texttt{and},
  \texttt{or}, or \texttt{xor}); and \texttt{r} and \texttt{v} indicate a register
  and a randomly chosen integer, respectively.}
\end{figure}

We extend \Disp{} in two main ways. First, we allow it to
displace any set of consecutive instructions within a basic
block, not only ones that belong to gadgets. Second, instead
of replacing the original instructions with traps, we
replace them with \emph{semantic nops}---sets of
instructions that \emph{cumulatively} do not affect the
memory or register values and have no side
effects~\cite{Christo05Semantics}. These semantic nops get
jumped to immediately after the displaced code is done
executing.

While nops can be defined atomically (e.g., by a
\texttt{nop} instruction), initial failures to mislead
malware detection indicated that a rich semantic nop
language is needed for successful attacks. Such a language
enables the attack to search through a large set of
functionally equivalent programs to evade \dnn{}s.
Therefore, we developed a context-free grammar to create
diverse semantic nops (see \figref{fig:semnops}). At a high
level, a semantic nop is an atomic instruction; or an
invertible instruction that is followed by a semantic nop
and then by the inverse instruction (e.g., \texttt{push eax}
followed by a semantic nop and then by \texttt{pop eax}); or
two consecutive semantic nops. When the flags register's
value is saved (i.e., between \texttt{pushfd} and
\texttt{popfd} instructions), a semantic nop may contain
instructions that affect flags (e.g., add and then subtract
a value from a register); and when a register's value is
saved (i.e., between \texttt{push r} and \texttt{pop r}), a
semantic nop may contain instructions that affect the
register (e.g., decrement it by a random value). Using the
grammar for generating semantic nops, for example, one may
generate a semantic nop that stores the flags and
\texttt{ebx} registers on the stack (\texttt{pushfd};
\texttt{push ebx}), performs an operation that might affect
both registers (e.g., \texttt{add ebx, 0xff}), and then
restores the registers (\texttt{pop ebx}; \texttt{popfd}).

\begin{figure}[!th]
  \centering
  \begin{subfigure}{0.30\textwidth}
    \hspace{2.25mm}\includegraphics[width=\textwidth]{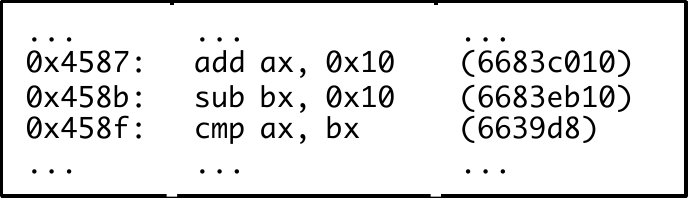}
    \caption{Original code\vspace{5pt}}
  \end{subfigure}

  \begin{subfigure}{0.325\textwidth}
    \includegraphics[width=\textwidth]{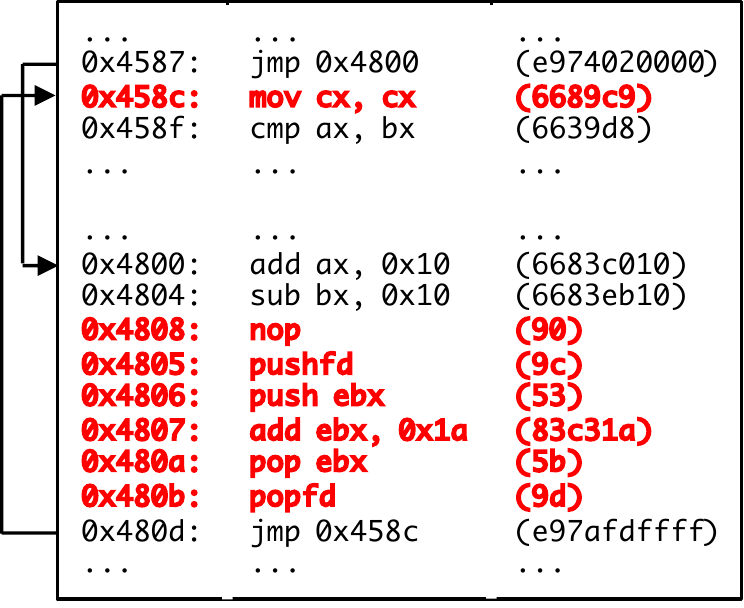}
    \caption{After \Disp{}}
  \end{subfigure}
  
  \caption{\label{fig:disp}An example of displacement. The two instructions
  staring at address \texttt{0x4587} in the original code (a) are displaced
  to starting address \texttt{0x4800}. The original instructions are replaced
  with a \texttt{jmp} instruction and a semantic nop. To consume the
  displacement budget, semantic nops are added immediately after the displaced
  instructions and just before the \texttt{jmp} that passes the control back to
  the original code. Semantic nops are shown in boldface and red.}
\end{figure}

When using \Disp{}, our attacks start by displacing code up
to a certain budget, to ensure that the resulting binary's
size does not increase above a threshold (e.g., 1\% above
the original size). We divide the budget (expressed as the
number of bytes to be displaced) by the number of functions
in the binary and attempt to displace exactly that number of
bytes per function. If multiple options exist for what code
in a function to displace, we choose at random. If a
function does not contain enough code to displace, then we
attach semantic nops after the displaced code to meet the
per-function budget. In the rare case that the function does
not have any basic block larger than five bytes, we skip
that function. \figref{fig:disp} illustrates an example of
displacement where semantic nops are inserted to replace
original code as well as after displaced code, to consume
the budget. Then, in each iteration of modifying the binary
to cause it to be misclassified, new semantic nops are
chosen at random and used to replace the previously inserted
semantic nops if that moves the binary closer to
misclassification.

Some of the semantic nops contain integer values that can be
set arbitrarily (e.g., see line 12 of \figref{fig:semnops}).
In a white-box setting, the bytes of the binary that
correspond to these values can be set to perturb the
embedding in the direction that is most similar to the
gradient. Namely, if an integer value in the semantic nop
corresponds to the $i$\textsuperscript{th} byte in the
binary, we set this $i$\textsuperscript{th} byte to
$b\in\{0,\ldots,255\}$ such that the cosine similarity
between $\embedFunc(b)-\embedFunc(\hat{x}_i)$ and $g_i$ is
maximized. This process is repeated each time a semantic nop
is drawn to replace previous semantic nops in white-box
attacks.

Known methods~\cite{Christodorescu05Norm} for detecting and
removing semantic nops from binaries might appear viable for
defending against \Disp{}-based attacks.  However, as we
discuss in \secref{sec:discussion}, attackers can leverage
various techniques to evade semantic-nop detection and
removal.

\parheading{Limitations} Our implementation leaves room for
improvement. For instance, it does not displace code that
has been displaced in earlier iterations. A better
implementation might apply displacements recursively.
Furthermore, the composability of \IPR{} and \Disp{}
transformations could be improved. In particular, when
applying both \Disp{} and \IPR{} transformations to a
binary, both types of transformations affect the original
instructions of the binary. However, \IPR{} does not affect
the semantic nops that are introduced by \Disp{}.  Despite
room for improvement, our implementation is already
sufficient to generate successful attacks (see below).

\section{Evaluation}
\label{sec:results}

In this section, we comprehensively evaluate our attack. We
first detail the \dnn{}s and data used for evaluation. We
then show that na\"ive, random transformations that are not
guided via optimization do not lead to misclassification.
Subsequently, we evaluate variants of our attack in the
white- and black-box setting and compare with prior work. We
then evaluate our attack against commercial anti-viruses and
close the section with experiments to validate that the
attacks preserve functionality.

\subsection{Datasets and Malware-Detection \dnn{}s}
\label{sec:res:data}

\begin{table}
\small
  \begin{center}
  \begin{tabular}  {r | c c c}
    \toprule
    \vtfeed{} & Train & Val. & Test \\ \midrule
    Benign & 111,258 & 13,961 & 13,926  \\
    Malicious & 111,395 & 13,870 & 13,906  \\
    \bottomrule
  \end{tabular}
  \caption{\label{tab:train-test-data} The number of benign and malicious
    binaries used to train, validate, and test the \dnn{}s.}
\end{center}
\end{table}

\subsubsection{Dataset composition}
Our dataset, \vtfeed{}, contains raw binaries of malware
samples targeting Windows machines. As such, the binaries
adhere to the Portable Executable format (\pe{}; the
standard format for \texttt{.dll} and \texttt{.exe}
files)~\cite{MS19PE}. Overall, we use significantly more
samples than similar prominent prior work
(e.g.,~\cite{Arp14Drebin, Kolter06Malware}).

\vtfeed{} was collected by sampling the VirusTotal feed for
PE binaries, representing binaries encountered in practice
by anti-virus vendors. Collection took around two weeks and
was restricted to binaries first seen in 2020, to ensure
recency, and smaller than 5 MB. Following prior
work~\cite{Anderson18Ember}, binaries were filtered and
labeled as benign (resp., malicious) if they were classified
as malicious by 0 (resp., over 40) antivirus vendors as
aggregated by VirusTotal. The dataset contains 278,316
binaries with a roughly even distribution between benign and
malicious binaries. We sampled training, test, and
validation sets at a ratio of 80\%, 10\%, and 10\%
respectively. Exact numbers can be seen in
Table~\ref{tab:train-test-data}.

\subsubsection{\dnn{} Training}
Using the malicious and benign samples, we trained two
malware-detection \dnn{}s. All \dnn{}s receive binaries' raw
bytes as inputs and output the probability that the binaries
are malicious. The first \dnn{} (henceforth, \avastnet{}),
proposed by Kr{\v{c}}{\'a}l et al.~\cite{Krvcal18AvastNet},
receives inputs up to 512~KB in size. The second \dnn{}
(henceforth, \malconv{}), proposed by Raff et
al.~\cite{Raff18MalConv}, receives inputs up to 2~MB in
size. Except for the batch size (set to 32 due to memory
limitations), we used the same training parameters reported
in prior work. When using binaries for training, we excluded
their headers so the \dnn{}s would not rely on header
values, which are easily manipulable, for
classification~\cite{Demetrio19FoolMD}.

\begin{table}
\small
  \begin{center}
  \begin{tabular}  {r | c c c | c}
    \toprule
           & \multicolumn{3}{c|}{Accuracy} & TPR @ \\
                & Train    & Val.    & Test    & 0.1\% FPR \\ \midrule
    \avastnet{} & 99.89\%  & 98.59\% & 98.60\% & 94.78\%  \\
    \malconv{}  & 99.97\%  & 98.67\% & 98.53\% & 96.08\% \\  
    \bottomrule
  \end{tabular}
  \caption{\label{tab:dnn-perf} The \dnn{}s' accuracy and
    the TPR at the operating point where the FPR equals
    0.1\%. }
\end{center}
\end{table}

Each \dnn{} achieves test accuracy of about 99\% (see
\tabref{tab:dnn-perf}). Even when restricting the false
positive rates (FPRs) conservatively to 0.1\% (as is often
done by anti-virus vendors~\cite{Krvcal18AvastNet}), the
true positive rates (TPRs) remain as high as 94--96\%
(i.e., 94--96\% of malicious binaries are detected). These
results are superior to those reported in the original
papers both for classification from raw bytes and from
manually crafted features~\cite{Krvcal18AvastNet,
Raff18MalConv}. This is likely because 
\vtfeed{} was sampled over a narrow time span, and expect
the performance would slightly decrease if we increased
the diversity of the dataset.

In addition to the two \dnn{}s that we trained, we evaluated
our attacks using a publicly available \dnn{} (henceforth,
\ember{}) trained by Anderson and
Roth~\cite{Anderson18Ember}. \ember{} has a similar
architecture to \malconv{}. The salient differences are
that: \emph{1)} \ember{}'s input dimensionality is 1 MB
(compared to 2 MB for \malconv{}); and \emph{2)} \ember{}
uses the \pe{} header for classification. On a dataset
curated by a computer-security company, \ember{} achieved
about 92\% TPR when the FPR was restricted to
0.1\%~\cite{Anderson18Ember}.

To evaluate attacks against the \dnn{}s, we selected
binaries according to three criteria. First, the binaries
had to be unpacked. We used standard packer detectors,
Packerid~\cite{Packerid} and Yara~\cite{Yara}, and deemed
binaries as unpacked only if no detector exhibited a
positive detection. This method is similar to the one
followed by Biondi et
al.~\cite{Biondi18PackDet}.\footnote{Biondi et al.\ used
three packer-detection tools instead of two. Unfortunately,
we were unable to get access to one of the proprietary
tools.} We also filtered out binaries labeled as packed in
their VirusTotal metadata. While the data used to train and
test the \dnn{}s included packed binaries, the high accuracy
of the \dnn{}s on the test samples suggests that the
\dnn{}s' performance was not impacted by (lack of) packing.
Second, the binaries had to be classified correctly and with
high confidence by the \dnn{}s that we trained. In
particular, malicious binaries had to be classified as
malicious and the estimated probability that they are
malicious had to be above the threshold where the FPR is
0.1\%. Consequently, our evaluation of the attacks' success
is conservative: the attacks would be more successful for
binaries that are initially classified correctly, but not
with high confidence. Third, the binaries' sizes had to be
smaller than the \dnn{}s' input dimensionality. We further
restricted the binaries' sizes to be smaller than smallest
input dimensionality of our \dnn{}s (\avastnet{} at 512~KB).
While the \dnn{}s can classify binaries whose size is larger
than the input dimensionality (as can be seen from the high
classification accuracy on the validation and test sets), we
avoided large binaries as a means to prevent evasion by
displacing malicious code outside the input range of the
\dnn{}s. Using these criteria, we selected 100 malicious
binaries from the test set to evaluate the attacks against
each of the three \dnn{}s.

The total number of samples we collected is comparable to
that used in prior work on evading malware
detection~\cite{Kolos18Malware, Kreuk18Malware,
Srndic14PDFfool, Suciu18Malware}.

\subsection{Attack-Success Criteria}
We executed the attacks for up to 200 iterations, stopping
early if the binaries were misclassified at the operating
point where the FPR equals 0.1\%. For malicious binaries,
this meant that they were misclassified as benign with a
probability higher than a model-specific threshold set to
achieve 0.1\% FPR. This follows the threshold typically used
by antivirus vendors (e.g.,~\cite{Krvcal18AvastNet}). We
also found that attack success on the same binary, given
identical experiment parameters, was often stochastic.
Therefore, we repeated each attack 10 times to get a
reliable measure of attack success.

We compared the overall success of attacks in two ways: by
the percentage of binaries that were misclassified in
\textit{at least} 1 of the 10 repeated attacks on them
(\textbf{\coverage{}}); and the overall percentage of
attacks that were successful across all attacked binaries
(\textbf{\potency{}}). Coverage\note{didn't use macro here
to capitalize word manually} is a measure of what percentage
of binaries our attack \textit{can} be successful on whereas
\potency{} is a measure of the how often a single attack
trial succeeds. As a result of this definition, \coverage{}
will always be higher than \potency{}.

\subsection{Randomly Applied Transformations}
\label{sec:results:rand}

We first evaluated whether na\"ively transforming binaries
at random would lead to evading the \dnn{}s. For each binary
that we used to evaluate the attacks we created 200 variants
using the \IPR{} and \Disp{} transformations and classified
them using the \dnn{}s. We transformed the binaries
sequentially and at random. Namely, starting from the
original variant, we created the next variant by
transforming every function using a randomly picked
transformation type that was applied at random. If any of
the variants were misclassified by a \dnn{} given the 1\%
FPR threshold, we would consider the evasion attempt
successful. We set \Disp{} to increase binaries' sizes by
5\% (i.e., the displacement budget was set to 5\% of the
binary's original size). We selected 200 and 5\% as
parameters for this experiment because our attacks were
executed for 200 iterations at most and achieved almost
perfect success when increasing binaries' sizes by 5\% (see
below). This technique was most effective when attempting to
misclassify malware as benign on \ember{}, where four
binaries evaded detection. However, for all other attempts
to evade, no more than three binaries were successful. 

Hence, we conclude the \dnn{}s are robust to na\"ive
transformations and more principled approaches are needed to
mislead them.

\subsection{White-Box Attacks vs.\ \dnn{}s}
\label{sec:results:attackwb}

\begin{figure*}[!ht]
  \centering
  \begin{subfigure}{\textwidth}
    \makebox[\textwidth][c]{\includegraphics[width=\textwidth]{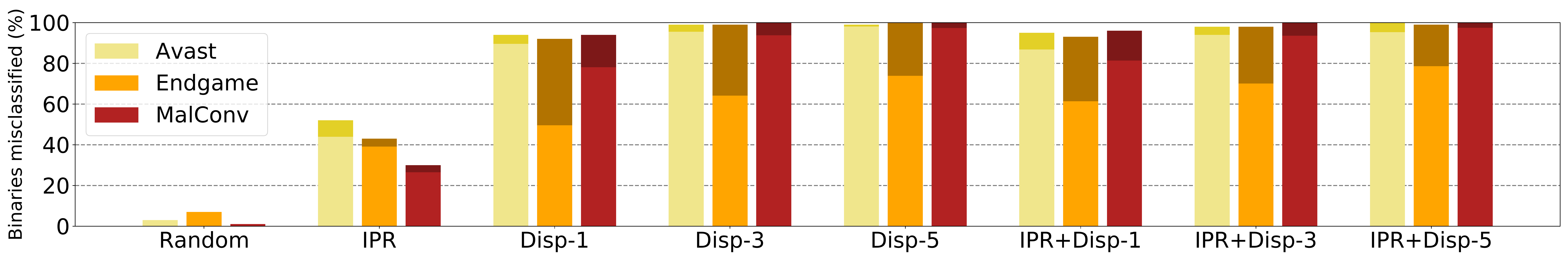}}
  \end{subfigure}

  \caption{\label{fig:wbattacks}Attack success rates in the
    white-box setting. We show \potency{} as the lighter
    bars and \coverage{} as the darker bars.}
\end{figure*}

In the white-box setting, we evaluated seven variants of our
attack. One variant, to which we refer to as \IPR{}, relies
on the \IPR{} transformations only. Three variants,
\Disp{}-1, \Disp{}-3, and \Disp{}-5, rely on the \Disp{}
transformations only, where the numbers indicate the
displacement budget as a percentage of the binaries' sizes
(e.g., \Disp{}-1 increases binaries' sizes by 1\%). The last
three attack variants, \IPR{}+\Disp{}-\{1,3,5\}, use the
\IPR{} and \Disp{} transformations combined.

We set 5\% as the maximum displacement budget and 200 as the
maximum number of iterations, as the attacks were almost
always successful with these parameters.

\begin{figure*}[!ht]
  \centering
  \begin{subfigure}{\textwidth}
    \makebox[\textwidth][c]{\includegraphics[width=.8\textwidth]{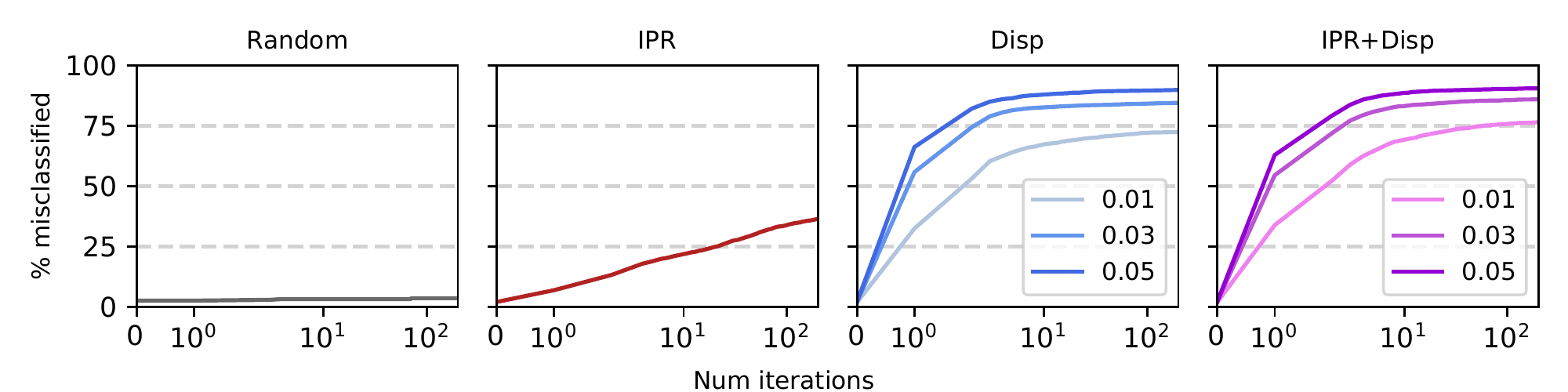}}
  \end{subfigure}
  \caption{\label{fig:wbattacktrendsbudget}A contrasting view showing the
    \potency{} over iteration for the whitebox attacks.} 
\end{figure*}

The results of the experiments are provided in
\figref{fig:wbattacks} where the lighter part of the bar
represents \potency{} and the darker part represents
\coverage{}. One can immediately see that attacks using the
\Disp{} transformations were more successful than \IPR{}.
While showing some effectiveness in evading \ember{}, \IPR{}
at best achieves a \coverage{} of 52\% while \Disp{} of all
budgets on all three models are able to cause at least 92\%
of binaries to be misclassified.

Moreover, \Disp{}-5 achieved misclassification on all
binaries except one on \avastnet{}. As one would expect,
attacks with higher displacement budgets were more
successful than attacks with lower displacement budgets.
However, the main difference we see is in the \potency{} of
the attack, whereas the \coverage{} only differs by a single
missed binary between \Disp{}-3 and \Disp{}-5.

In addition to achieving higher \coverage{} and \potency{},
another advantage of \Disp{}-based attacks over \IPR{}-based
ones is their time efficiency. While displacing instructions
at random from within a function with $n$ instructions has
$\mathcal{O}(n)$ time complexity, certain \IPR{}
transformations have $\mathcal{O}(n^2)$ time complexity. For
example, reordering instructions requires building a
dependence graph and extracting instructions one after the
other. If every instruction in a function depends on
previous ones, this process takes $\mathcal{O}(n^2)$ time.
In practice, we found that \IPR{}--based, \Disp{}--based,
and \IPR{}+\Disp{}--based attacks took 4424, 283, and
961 seconds on average respectively.\footnote{Times
were computed on four machines: one with 2.2GHz AMD Opteron
CPU and 128GB RAM, one with 3.4GHz Intel-i7 CPU and
24GB RAM, one with 2.2Ghz AMD Ryzen 3900X and 64GB
RAM, and one with 2.7GHz Intel-i5 CPU and 24GB
RAM.}

Combining \IPR{} with \Disp{} achieved noticeably better
results in fewer iterations than respective \Disp{}-only
attacks when the budget for \Disp{} is low. For example,
\IPR{}+\Disp{}-1 had 11\% higher \potency{} than \Disp{}-1
when misleading \ember{} to misclassify a malicious binary
as benign (61\% vs.\ 50\% \potency{}). Thus, in certain
situations, \Disp{} and \IPR{} can be combined to fool the
\dnn{}s while increasing binaries' sizes less than \Disp{}
alone.

For our most performant attack, \IPR{}+\Disp{}-5, we
re-executed the attacks with significantly more difficult
success criteria. We changed the threshold for attack
success to \malconv{} and \avastnet{}'s FNR of 0.01\%.
Beating this threshold means that a transformed binary must
appear less malicious than the least malicious 0.1\% of
malware in the dataset. For \malconv{}, our \potency{} drops
from 97\% to 92\%, while \coverage{} drops from 100\% to
99\%. For \avastnet{}, \potency{} drops from 95\% to 90\%
and \coverage{} from 100\% to 95\%. These results
demonstrate our attack's ability to evade more cautious ML
detectors, even though this threshold is unlikely to be used
as it would flag roughly a third of benign binaries as
malware.

In \figref{fig:wbattacktrendsbudget}, we averaged and
plotted the classification output of the models and the
resultant misclassifications of the binaries over the
iterations of each attack. As shown, the majority of
successful attacks that incorporate \Disp{} succeeded in a
single iteration, with almost all successful attacks
occurring within ten iterations.
\begin{figure*}[!ht]
  \centering
  \begin{subfigure}{0.23\textwidth}
    \includegraphics[width=1.0\textwidth]{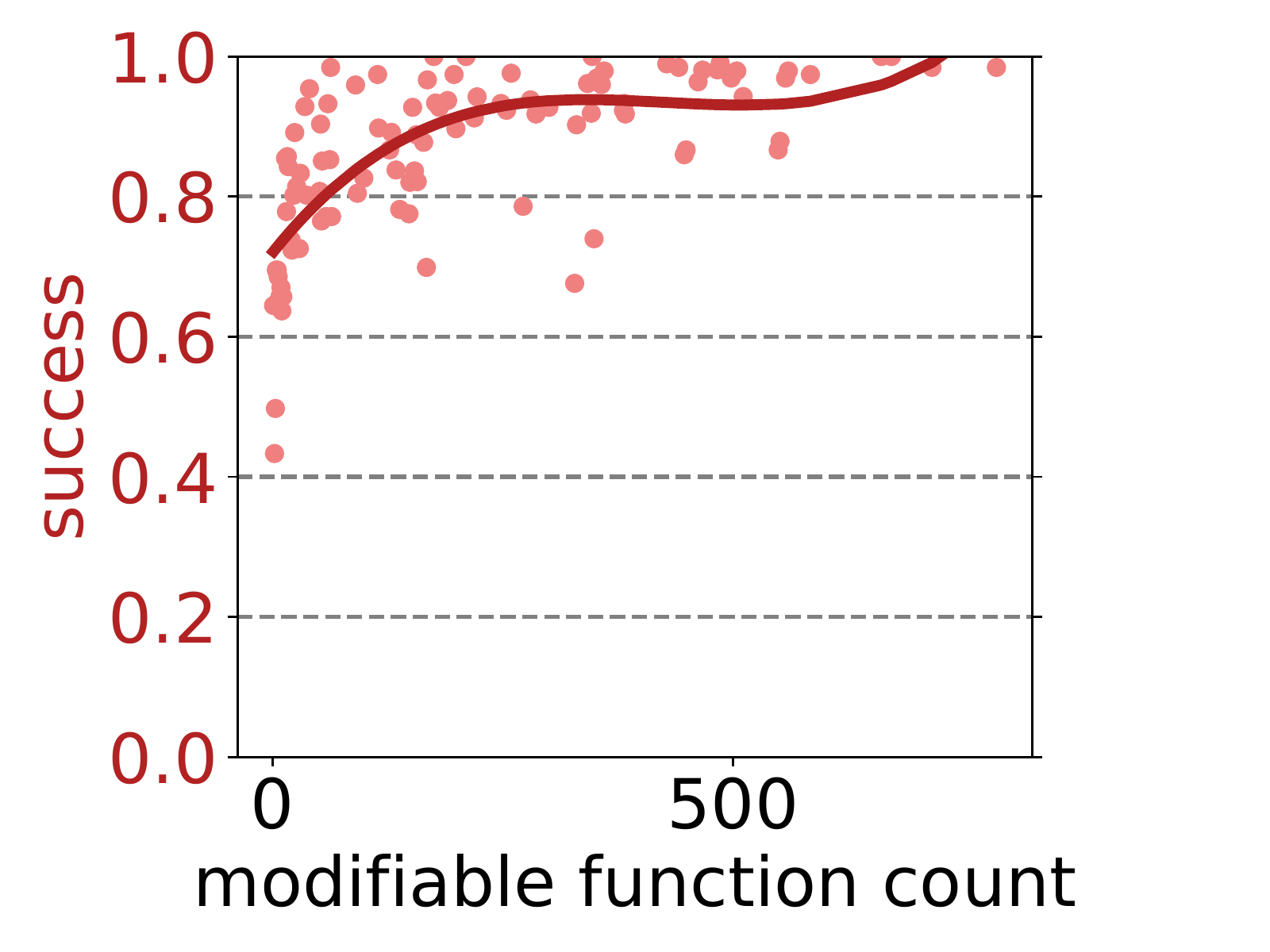}
    \caption{\label{fig:binpropstats:a}Success (all attacks)}
  \end{subfigure}\hspace{2mm} 
  \begin{subfigure}{0.23\textwidth}
    \includegraphics[width=1.0\textwidth]{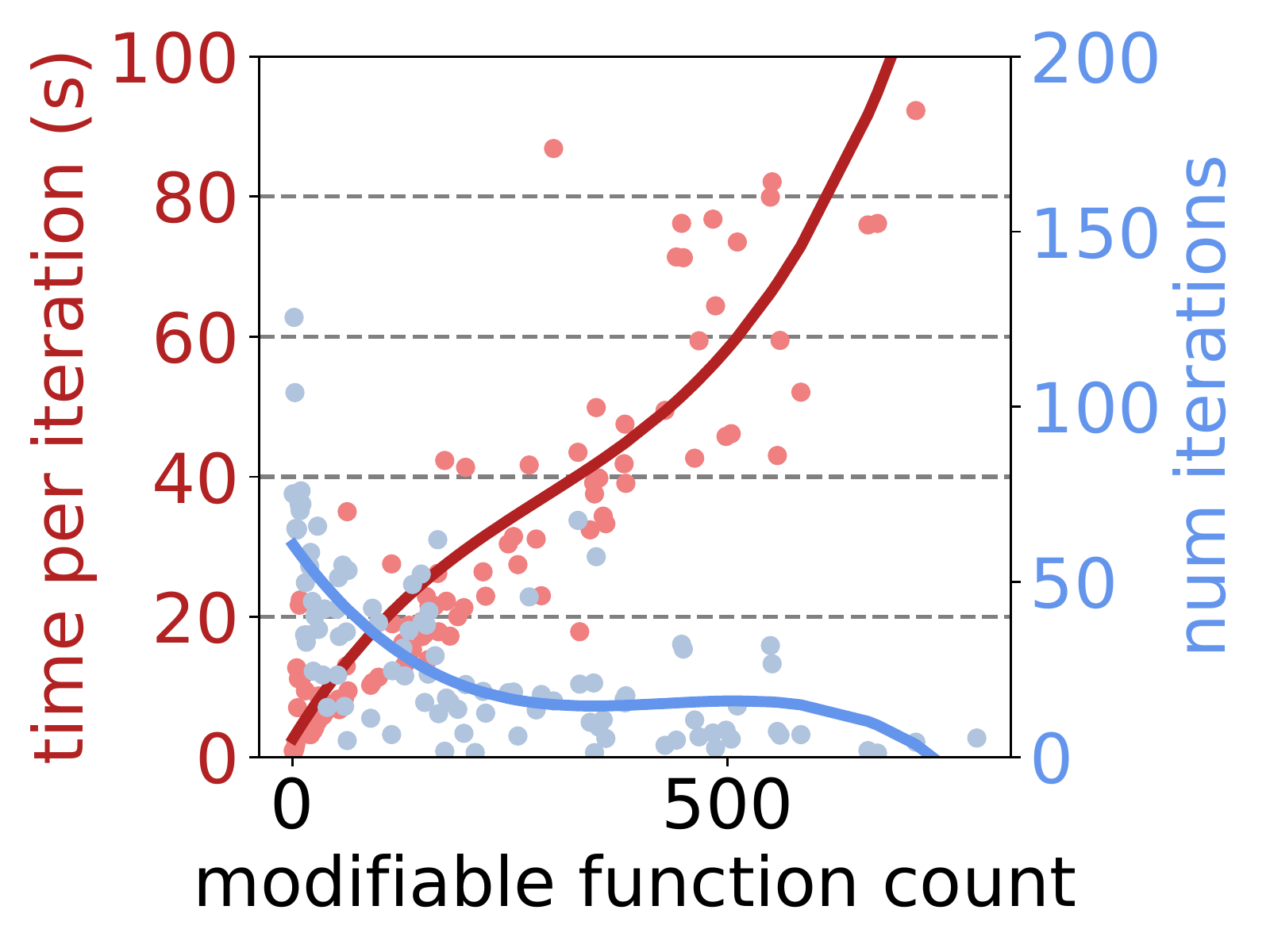}
    \caption{\label{fig:binpropstats:b}Time (all attacks)}
  \end{subfigure}\hspace{2mm} 
  \begin{subfigure}{0.23\textwidth}
    \includegraphics[width=1.0\textwidth, valign=t]{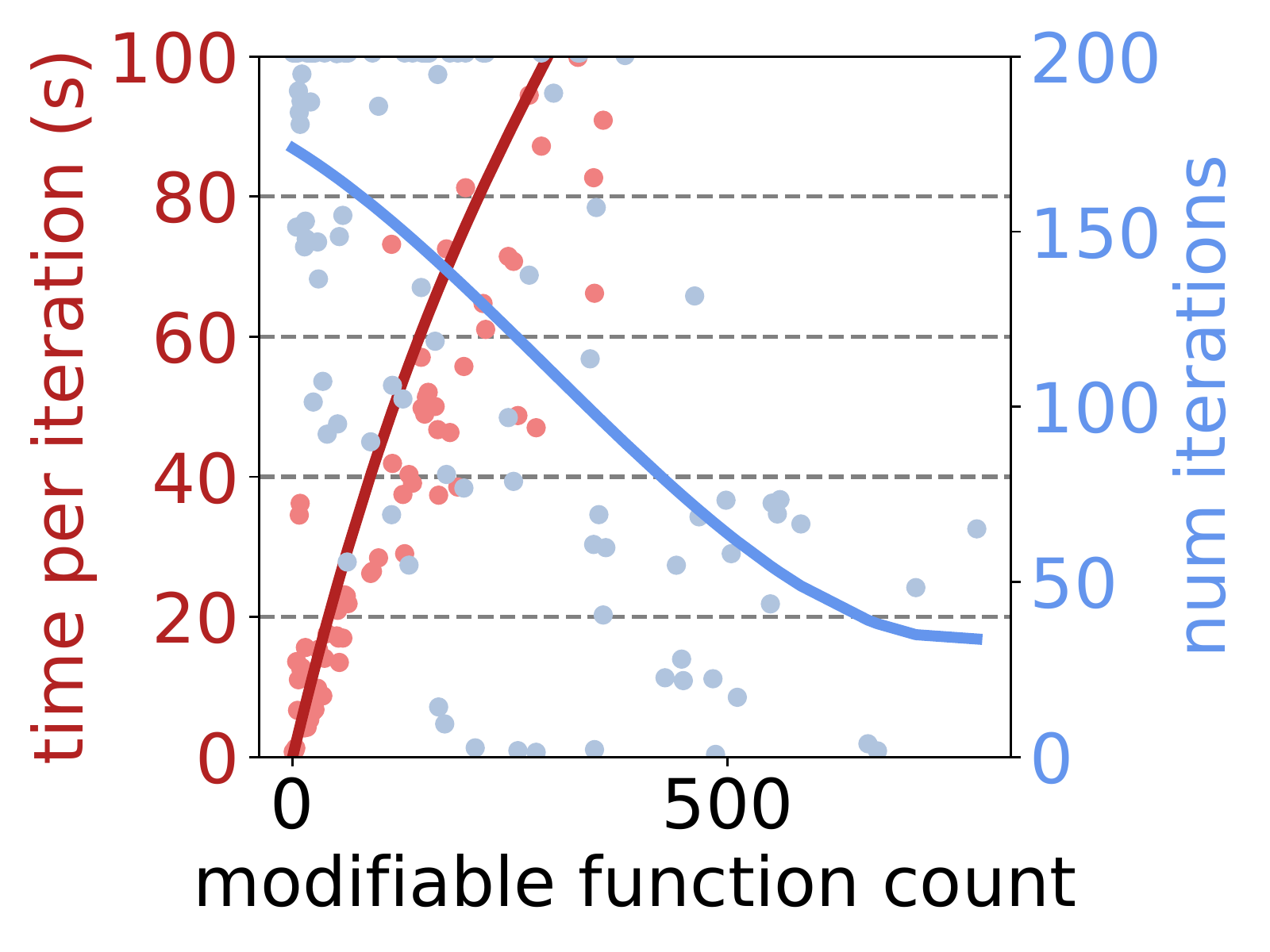}
    \caption{\label{fig:binpropstats:c}Time (\IPR{})}
  \end{subfigure}\hspace{2mm} 
  \begin{subfigure}{0.23\textwidth}
    \includegraphics[valign=t,width=1.0\textwidth]{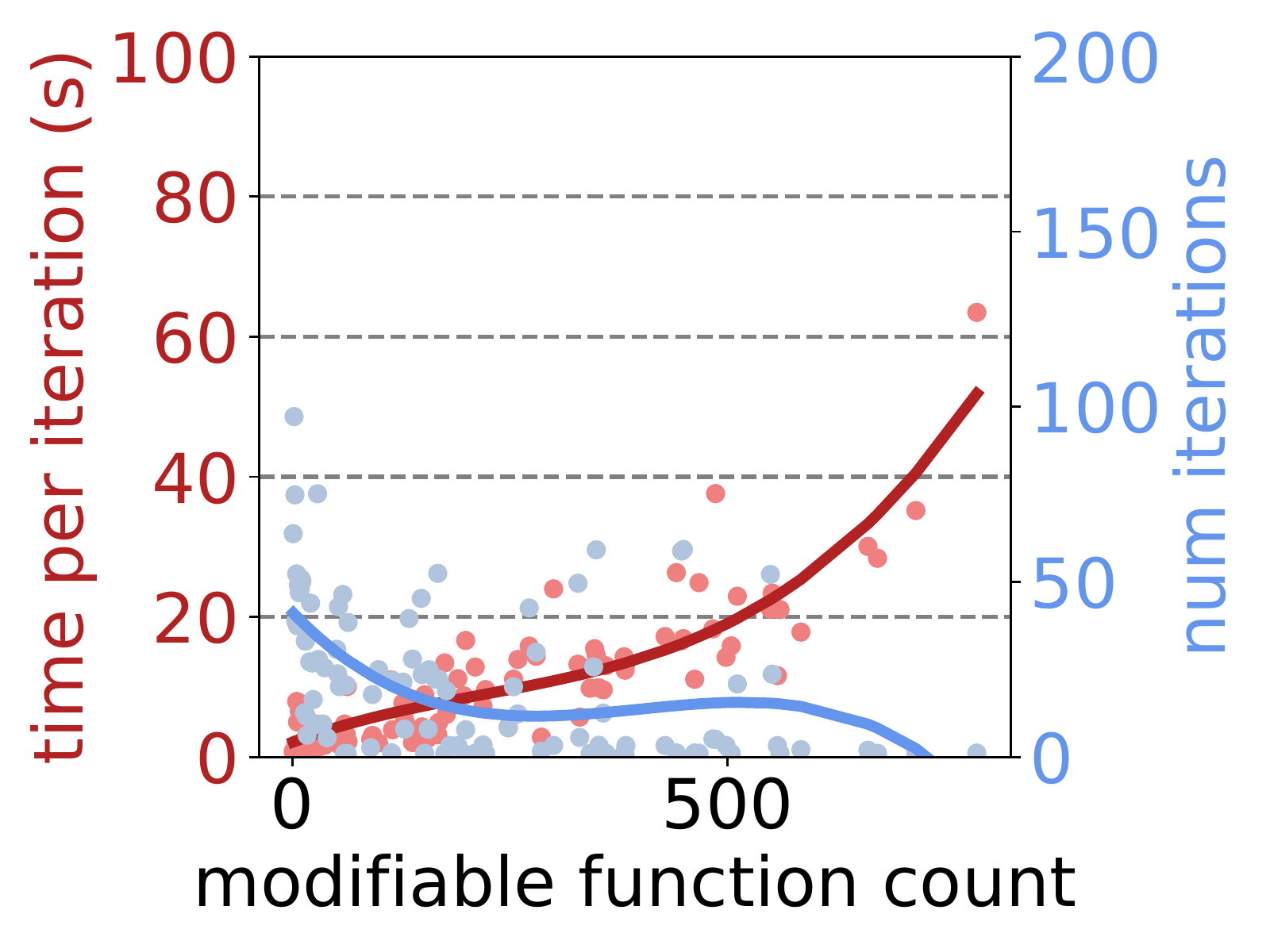}
    \caption{\label{fig:binpropstats:d}Time (\Disp{})}
  \end{subfigure}

  \caption{\label{fig:binpropstats} As the number of
  modifiable functions increased, the average number of
  iterations to success decreased, while the time to execute
  an iteration increased. The lines in each plot are the
  best fit degree-3 polynomials.}
\end{figure*}
 We also
examined the performance of the attacks as a function of the
number of modifiable functions in a binary. On average, 89\%
of functions in a binary were modifiable. As expected, attacks were less likely
to succeed when the binaries had few functions to modify
(\figref{fig:binpropstats:a}). Consistent with that finding,
as the number of modifiable functions (and number of functions
overall) in a binary increased, the average number of
iterations required for an attack to succeed decreased
(\figrefs{fig:binpropstats:b}{fig:binpropstats:d}). This
trend held across different types of attacks, but was more
pronounced for less successful attacks (\IPR{}) than more successful ones
(\Disp{}), as the vast majority of the latter completed within a small number of
iterations. 

Finally, we compared the evasion success rates of our attack
with a representative prior attack proposed by Kreuk et
al.~\cite{Kreuk18Malware}. 
To mislead \dnn{}s, this attack appends
adversarially crafted bytes to binaries. These bytes are
crafted via an iterative algorithm that first computes the
gradient $g_i$ of the loss with respect to the embedding
$\embedFunc(x_i)$ of the binary $x_i$ at the
$i$\textsuperscript{th} iteration, and then sets the
adversarial bytes to minimize the $L_2$ distance of the new
embedding $\embedFunc(x_{i+1})$ from
$\embedFunc(x_i)$+$\epsilon\mathit{sign}(g_i)$, where
$\epsilon$ is a scaling parameter. We tested three variants
of the attack which increase the binaries' sizes by 1\%,
3\%, and 5\%. We used $\cwloss$ as the loss function. As
with our attacks, we executed Kreuk et al.'s attacks for up
to 200 iterations, stopping sooner if misclassification
occurred. We set $\epsilon$=1, as we empirically found it
leads to high evasion success.

The variants of Kreuk et al.'s attack achieved success rates
comparable to our attack. \Kreuk{}-5 was almost always able
to mislead the \dnn{}s---it achieved 99\% and 98\% success
rate when attempting to mislead \ember{} and \malconv{},
respectively, to misclassify malicious binaries, and 100\%
success rate in all other attempts. Also similar to our
attacks, the success rates increased as the attacks
increased the binaries' sizes. However, as described in
\appref{app:sanitization}, their attack is easier to defend
against by sanitizing bytes (specifically, by masking with
zeros) in sections that do not contain instructions.

\subsection{Black-Box Attacks vs.\ \dnn{}s}

As explained in \secref{sec:tech_approach}, because the
\dnn{}s' input is discrete, estimating gradient information
to mislead them in a black-box setting is not possible. So,
the black-box version of \algref{alg:attack} uses hill
climbing to query the \dnn{} after each attempted
transformation to decide whether to keep the transformation.
Because querying the \dnn{}s after each attempted
transformation significantly increased the run time of the
attacks ($\sim$30$\times$ on a machine with GeForce GTX 980
GPU), we limited our experiments to \Disp{} transformations
with a displacement budget of 5\%. We executed the attacks
up to 200 iterations, stopped early if misclassification
occurred, and repeated them three times each to account for
stochasticity.

The attacks were most successful against \malconv{},
achieving a \coverage{} of 95\% and \potency{} of 92\%.
\avastnet{} and \ember{} were only slightly more robust with
attack \coverage{}s of 92\% and 59\% and potencies of 87\%
and 56\% respectively. These results show our attack remains
effective even in a black-box setting.

\subsection{Commercial Anti-Viruses}
\label{sec:vttransfer}

To assess whether our attacks affect commercial
anti-viruses, we tested the malicious transformed binaries
that were misclassified by the \dnn{}s in the white-box
setting on the anti-viruses available via
VirusTotal~\cite{VirusTotal}---a service that aggregates the
results of 68 commercial anti-viruses. Since anti-viruses
often rely in part on static analysis, with increasing
integration of \ml{}, we expected that the malicious
binaries generated by our attacks would be detected by fewer
anti-viruses than the original binaries.

Due to contractual constraints, we were unable to perform
this experiment with our previously described dataset. Thus,
we resorted to using binaries taken from other sources. In
this alternate dataset, we used 21,741 malicious binaries
belonging to seven malware families that were published by
Microsoft as part of a malware-classification
competition~\cite{Ronen18MSRKaggle}. We complemented these
binaries with 19,534 benign binaries collected by installing
standard packages (browsers, productivity tools, etc.) on a
newly created 32-bit Windows 7 virtual
machine.\footnote{Specifically, we used the Ninite and
Chocolatey (\url{https://ninite.com/} and
\url{https://chocolatey.org/}) package managers to install
179 packages.} After splitting the binaries for training
(21,217), testing (9,105), and validation (10,953), we
trained variants of \malconv{} and \avastnet{} that achieved
99.15\% and 98.92\% test accuracy, respectively.
Subsequently, we collected 95 malicious binaries from
VirusShare~\cite{VirusShare} that pertain to the seven
malware families from the Microsoft competition. We then
transformed these malicious binaries using our white-box
attack to evade the \dnn{}s we trained as well as \ember{},
and tested how often the transformed binaries were detected
by anti-viruses on VirusTotal.

\parheading{Original Binaries}
As a baseline, we first classified the original binaries
using the VirusTotal anti-viruses. As one would expect, all
the malicious binaries were detected by several
anti-viruses. The median number of anti-viruses that
detected any particular malware binary as malicious was 55,
out of 68 total anti-viruses.

\parheading{Random Transformations}
To further gauge the efficacy of our guided attack over
random diversification, we used commercial anti-viruses to
classify binaries that were transformed at random using the
\Disp{} and \IPR{} transformations (as described in
\secref{sec:results:rand}). We found that certain
anti-viruses were susceptible to such simple evasion
attempts, presumably due to using fragile detection
mechanisms such as signatures. The median number of
anti-viruses that correctly detected the malicious binaries
decreased from 55 to 43.

\parheading{Packing}
We tested whether anti-viruses were susceptible to evasion
via packing. We used UPX~\cite{UPX}, one of the most popular
packers~\cite{Roundy13Packing}, and packed binaries
using the highest compression ratios. Interestingly, packing
malicious binaries was counter-productive for evading
anti-viruses. Packed malicious binaries were more likely to
be detected as malware---the median number of anti-viruses
that correctly detected malicious binaries increased from 55
for the original binaries to 59 after packing. 

\parheading{Our Attacks}
Compared to the original malicious binaries and randomly
transformed ones, the malicious binaries transformed by our
attacks were detected by fewer anti-viruses.
The median number of anti-viruses that correctly detected
the malicious binaries decreased from 55 for the original
binaries and 42 for ones transformed at random to 33--36,
depending on the attack variant and the targeted \dnn{}.
According to a Kruskal-Wallis test, this reduction is
statistically significant ($p<$0.01 after Bonferroni
correction). In other words, the malicious binaries that
were transformed by our attacks were detected by only
49\%--53\% of the VirusTotal anti-viruses in the median
case. \tabref{tab:vttransfer} in \appref{app:vttransfer}
summarizes each attack variant's effect on the number of
positive detections by anti-viruses.

Because our attack should not affect any dynamic analysis
(due to the desired attack property of functional
invariance), these results indicate some anti-viruses may be
over-reliant on static analyses and/or \ml{}. We also
highlight these results cannot only be attributed to
breaking signature-based defenses, as the randomly
transformed binaries (which were transformed for an equal
number of iterations) would have been equally likely to
evade anti-viruses as our attacks.

Furthermore, several anti-virus vendors that were misled by
our attacks advertise the use of \ml{} detectors. Evading
the \ml{} detectors of those vendors was necessary to
mislead their anti-viruses. A glance at vendors' websites
showed that 15 of the 68 vendors explicitly advertise
relying on \ml{} for malware detection. These anti-viruses
were especially susceptible to evasion by our attacks. Even
more concerning, a popular and highly credible anti-virus
whose vendor claims to rely on \ml{} misclassified 85\% of
the malicious binaries produced by one of our attacks as
benign. Generally, malicious binaries that were produced by
our attacks were detected by a median number of 7--9
anti-viruses of the 15---down from 12 positive detections
for the original binaries. All in all, our results support
that binaries that were produced by our attacks were able to
evade \ml{}-based static detectors that are used by
anti-virus vendors.

\subsection{Correctness}
\label{sec:results:correct}

A key feature of our attacks is that they transform binaries
to mislead \dnn{}s while preserving their functionality. We
followed standard practices from the binary-diversification
literature~\cite{Koo18Rand, Koo16Rand, Pappas12Rand} to
ensure that the functionality of the binaries was kept
intact after being processed by our attacks. First, we
transformed ten different benign binaries (e.g.,
\texttt{python.exe} of Python version 2.7, and
Cygwin's\footnote{\url{https://www.cygwin.com/}}
\texttt{less.exe} and \texttt{grep.exe}) with our attacks
and manually validated that they functioned properly after
being transformed. For example, we were still able to search
files with \texttt{grep} after the transformations. Second,
we transformed the \texttt{.exe} and \texttt{.dll} files of
a stress-testing
tool\footnote{\url{https://www.passmark.com/products/performancetest/}}
with our attacks and checked that the tool's tests passed
after the transformations. Using stress-testing tools to
evaluate binary-transformation correctness is common, as
such tools are expected to cover most branches affected by
the transformations. Third, and last, we also transformed
ten malware binaries and used the Cuckoo
Sandbox~\cite{Guarnieri12Cuckoo}---a popular sandbox for
malware analysis---to check that their behavior remained the
same. All ten binaries attempted to access the same hosts,
IP addresses, files, APIs, and registry keys before and
after being transformed.

\section{Potential Mitigations}
\label{sec:discussion}

Our proposed attacks achieved high success rates at fooling
\dnn{}s for malware detection in white-box and black-box
settings. The attacks were also able to mislead commercial
anti-viruses, especially ones that leverage \ml{}
algorithms. To protect users and their systems, it is
important to develop mitigation measures to make malware
detection robust against evasion by our attacks.  Our
efforts to explore mitigations, however, have met with
limited success.

Due to space limitations, we defer a complete discussion of
mitigations we have explored to \appref{app:mitigations},
providing only a brief summary here.  We found adversarial
training (e.g.,~\cite{Gdfllw14ExpAdv, Kurakin16AdvTrain})
too expensive in this domain to be practical.  Leveraging
static or dynamic analysis methods to ``cleanse'' binaries
is challenging because our attacks transform binaries'
original code (vs.\ inserting unreachable code) and can be
confounded through the insertion of
opaque~\cite{Collberg97Opaque, Moser07Opaque} or
evasive~\cite{Barak14Evasive} predicates. We found masking
random subsets of the binary prior to classification
promising as a defense in some cases, but it is far from
comprehensive and likely a small obstacle to an adaptive
attacker. Finally, detecting adversarial samples based on
binary size or \texttt{jmp} instructions seems both
difficult (our attacks increase neither substantially) and
ultimately evadable.  As further discussed in
\appref{app:mitigations}, we thus advocate that ML-based
static malware detection be augmented with methods not based
on ML.

\section{Conclusion}
\label{sec:conclusion}

Our work proposes evasion attacks on \dnn{}s for malware
detection. Differently from prior work, the attacks do not
merely insert adversarially crafted bytes to mislead
detection. Instead, guided by optimization processes, our
attacks transform the instructions of binaries to fool
malware detection while keeping functionality of the
binaries intact. As a result, these attacks are challenging
to defend against. We conservatively evaluated different
variants of our attack against three \dnn{}s under white-box
and black-box settings, and found the attacks successful as
often as 100\% of the time. Moreover, we found that the
attacks pose a security risk to commercial anti-viruses,
particularly ones using \ml{}, achieving evasion success
rates of up to 85\%. We explored several potential defenses,
and found some to be promising. Nevertheless, adaptive
adversaries remain a risk, and we recommend the deployment
of multiple detection algorithms, including ones not based
on \ml{}, to raise the bar against such adversaries.

\ifpublishable
\section*{Acknowledgments}

We would like to thank Leyla Bilge, Sandeep Bhatkar, Yufei
Han, and Kevin Roundy for helpful discussions. This work was
supported in part by the Multidisciplinary University
Research Initiative (MURI) Cyber Deception grant under ARO
award W911NF-17-1-0370; by NSF grants 1801391 and 2113345;
by the National Security Agency under award H9823018D0008;
by gifts from Google and Nvidia, and from Lockheed Martin
and NATO through Carnegie Mellon CyLab; by a CyLab
Presidential Fellowship and a NortonLifeLock Research Group
Fellowship; and by a DoD National Defense Science and
Engineering Graduate fellowship.
\fi

\bibliographystyle{ACM-Reference-Format}
\bibliography{cited}

\clearpage

\appendix
\section{Comparison to Kreuk et al. and Success After Sanitization}
\label{app:sanitization}

\begin{figure}[!ht]
  \centering
    \includegraphics[height=1.2in]{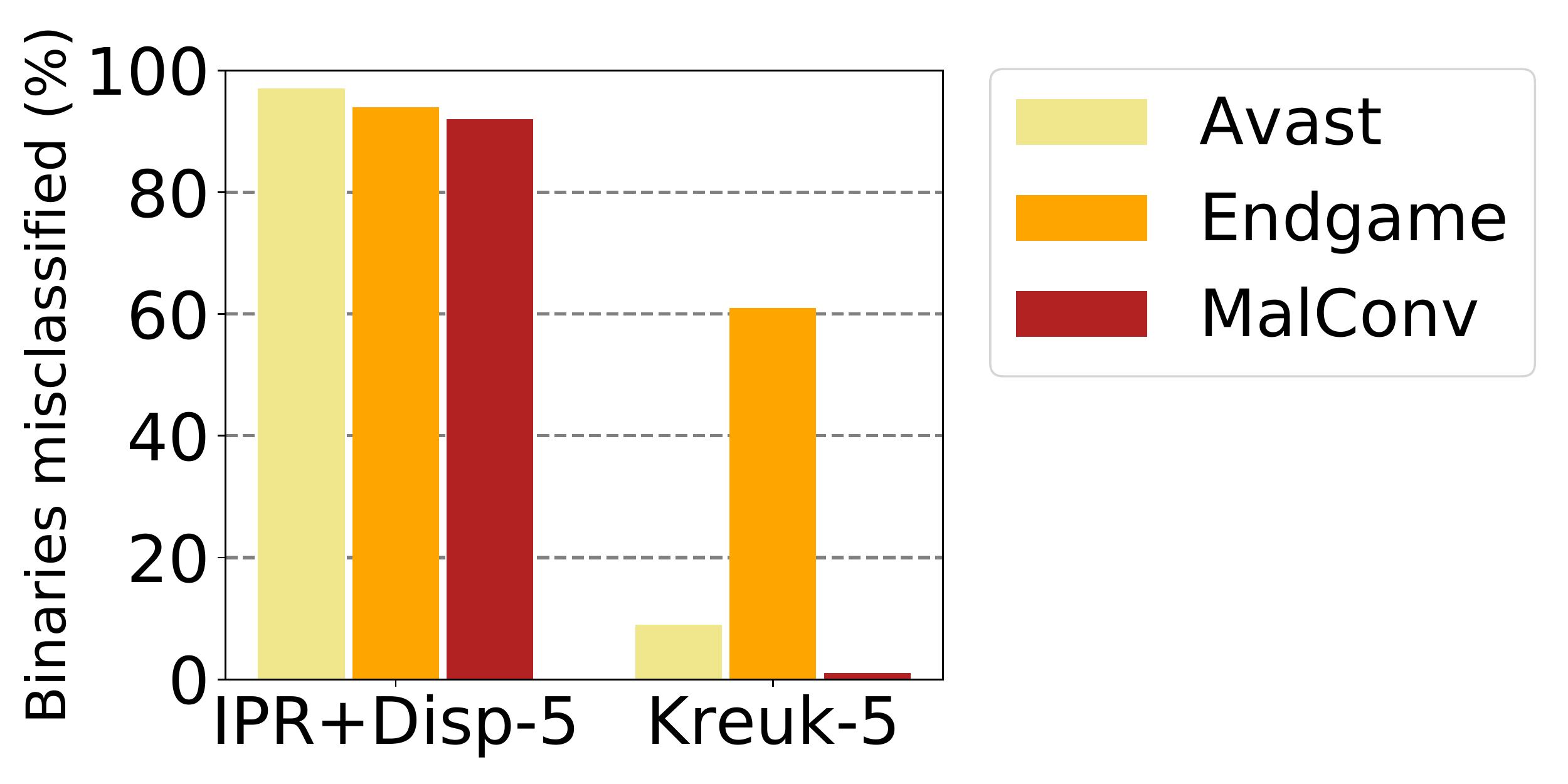}

    \caption{\label{fig:sanitization}Attacks' success rates (measured by the
      percentage of misclassified binaries) in the white-box setting when
      masking out bytes in sections that do not include instructions before
      classification.
    }
\end{figure}

While Kreuk et al.'s attack achieved success rates
comparable to ours, their attack is easier to defend
against. As a proof of concept, we implemented a
sanitization method to defend against the attack using our
alternate dataset described in \secref{sec:vttransfer}. The
method finds all the sections in a binary that do not
contain instructions (using the IDAPro
disassembler~\cite{IDAPro}) and masks the sections' content
with zeros. As Kreuk et al.'s attack does not add functional
instructions to the binaries, the defense masks the
adversarial bytes that the attack introduces. Consequently,
the evasion success rates of the attack drop significantly.
In fact, except for when attempting to mislead the \ember{}
\dnn{} with malicious binaries, the success rates of the
\Kreuk{} attacks dropped below 15\%. This defense had
little-to-no effect on our attacks, however: e.g., \Disp{}-5
still achieved 92\% and 100\% success rates against
\malconv{} for malicious and benign binaries, respectively.
Moreover, the classification accuracy remained high both for
malicious (99\%) and benign (93\%) binaries after the
defense. \figref{fig:sanitization} in
\appref{app:sanitization} presents the full results of the
impact of sanitization on attacks' success on the \kaggle{}
dataset.

\figref{fig:sanitization} shows the success rates of attacks
when sanitizing bytes in sections that do not include
instructions. In particular, we replaced byte values in such
sections with zeros, as described in
\secref{sec:results:attackwb}. Our attacks maintained high
success rates after sanitization (e.g., $>$90\% for
\Disp{}-5), whereas the success rates of the \Kreuk{}
attacks dropped below 15\% in most cases.

\section{Our Attacks' Transferability to Commercial Anti-Viruses}
\label{app:vttransfer}

\tabref{tab:vttransfer} summarizes the effect of different attack
variants on the number of positive detections (i.e., classification
of binaries as malicious) by the anti-viruses featured on VirusTotal.
\secref{sec:vttransfer} describes the experiment and explains the
results.

\begin{table*}
  \small
  \centering
  
\begin{subtable}{1\textwidth}
\centering
\begin{tabular}{r | c c c c c c c}
  \toprule
  \dnn{} & \IPR{} & \Disp{}-1 & \Disp{}-3 & \Disp{}-5 &
    \IPR{}+\Disp{}-1 & \IPR{}+\Disp{}-3 & \IPR{}+\Disp{}-5 \\ \midrule
    \avastnet{} & - & \textbf{36} & \textbf{35} & \textbf{36} & \textbf{36} & \textbf{35} & \textbf{36} \\
    \ember{} & \textbf{33} & \textbf{35} & \textbf{36} & \textbf{35} & \textbf{35} & \textbf{36} & \textbf{35} \\
    \malconv{} & - & \textbf{36} & \textbf{35} & \textbf{36} & \textbf{36} & \textbf{35} & \textbf{36} \\
  \bottomrule
\end{tabular}
\end{subtable}

\caption{\label{tab:vttransfer}The median number of
  VirusTotal anti-viruses that positively detected (i.e., as
  malicious) malicious binaries that were transformed by our
  white-box attacks (columns) to mislead the different
  \dnn{}s (rows). The median number of anti-viruses that
  positively detected for the original malicious binaries is
  55. Cases in which the change in the number of detections
  is statistically significant are in bold.}
\end{table*}

\section{Potential Mitigations}
\label{app:mitigations}

In this appendix we summarize our efforts to provide mitigations
for our attacks.

\subsection{Prior Defenses}
We considered several prior defenses to mitigate our
attacks, but, unfortunately, most showed little promise. For
instance, adversarial training (e.g.,~\cite{Gdfllw14ExpAdv,
Kurakin16AdvTrain}) is currently infeasible, as the attacks
are computationally expensive. Depending on the attack
variant, it took an average of 283 to 4424 seconds to run an
attack. As a result, running just a single epoch of
adversarial training would to take several weeks (using our
hardware configuration), as each iteration of training
requires running an attack for every sample in the training
batch. Moreover, while adversarial training might increase
the \dnn{}s' robustness against attackers using certain
transformation types, attackers using new transformation
types may still succeed at
evasion~\cite{Engstrom17AdvTrans}. Defenses that provide
formal guarantees
(e.g.,~\cite{Kolter17Defense,Mirman18Defense}) are even more
computationally expensive than adversarial training.
Moreover, those defenses are restricted to adversarial
perturbations that, unlike the ones produced by our attacks,
have small \lpnorm{\infty}- and \lpnorm{2}-norms. Prior
defenses that transform the input before classification
(e.g., via quantization~\cite{Xu18Squeeze}) are designed
mainly for images and do not directly apply to binaries.
Lastly, signature-based malware detection would not be
effective, as our attacks are stochastic and produce
different variants of the binaries after different
executions.

Differently from prior attacks on \dnn{}s for malware
detection~\cite{Kolos18Malware, Kreuk18Malware,
Suciu18Malware}, our attacks do not merely append
adversarially crafted bytes to binaries, or insert them
between sections. Such attacks may be defended against by
detecting and sanitizing the inserted bytes via static
analysis methods (e.g., similarly to the proof of concept
shown in \secref{sec:results:attackwb}, or using other
methods~\cite{Kruegel04Disas}). Instead, our attacks
transform binaries' original code, and extend binaries only
by inserting instructions that are executed at run time at
various parts of the binaries. As a result, our attacks are
difficult to defend against via static or dynamic analyses
methods (e.g., by detecting and removing unreachable code),
especially when augmented by measures to evade these
methods.

Binary normalization~\cite{Armoun12Norm,
Christodorescu05Norm, Walenstein06Norm} is a defense that
initially seemed viable for defending against our attacks.
The high-level idea of normalization is to employ certain
transformations to map binaries to a standard form and thus
undo attackers' evasion attempts before classifying the
binaries as malicious or benign. For example, Christodorescu
et al.\ proposed a method to detect and remove semantic nops
from binaries before classification, and showed that it
improves the performance of commercial
anti-viruses~\cite{Christodorescu05Norm}. To mitigate our
\Disp{}-based attacks, we considered using the semantic nop
detection and removal method followed by a method to restore
the displaced code to its original location. Unfortunately,
we realized that such a defense can be undermined using
\emph{opaque predicates}~\cite{Collberg97Opaque,
Moser07Opaque}. Opaque predicates are predicates whose value
(w.l.g., assume true) is known a priori to the attacker, but
is hard for the defender to deduce. Often, they are based on
$\mathit{NP}$-hard problems~\cite{Moser07Opaque}. Using
opaque predicates, attackers can produce semantic nops that
include instructions that affect the memory and registers
only if an opaque predicate evaluates to false. Since opaque
predicates are hard for defenders to deduce, the defenders
are likely to have to assume that the semantic nops impact
the behavior of the program. As a result, the semantic nops
would survive the defenders' detection and removal attempts.
As an alternative to opaque predicates, attackers can also
use \emph{evasive predicates}---predicates that evaluate to
true or false with an overwhelming probability (e.g.,
checking if a randomly drawn 32-bit integer is equal to
0)~\cite{Barak14Evasive}. In this case, the binary will
function properly the majority of the time, and may function
differently or crash once every many executions.

The normalization methods proposed by prior work would not
apply to the transformations performed by our \IPR{}-based
attacks. Therefore, we explored methods to normalize
binaries to a standard form to undo the effects of \IPR{}
before classification. We found that a normalization process
that leverages the \IPR{} transformations to produce the
form with the lowest lexicographic representation (where the
alphabet contains all possible 256 byte values) prevented
\IPR{}-based attacks. Formally, if $[x]$ is the equivalence
class of binaries that are functionally equivalent to $x$
and that can be produced via the \IPR{} transformation
types, then the normalization process produces an output
$\normFunc(x)\in [x]$, such that, $\normFunc(x)\le x_i$ for
every $x_i\in [x]$. \appref{app:iprnorm} presents an
algorithm that computes the normalized form of a binary when
executed for a large number of iterations, and approximates
it when executed for a few iterations. At a high level, the
algorithm applies the \IPR{} transformations iteratively in
an effort to reduce the lexicographic representation after
every iteration. We found that executing the algorithm for
ten iterations was sufficient to defend against \IPR{}-based
attacks. In particular, we executed the normalization
algorithm using the malicious and benign binaries produced
by the \IPR{}-based attacks to fool \ember{} in the
white-box setting, and found that the success rates dropped
to 3\% and 0\%, respectively, compared to 62\% and 74\%
before normalization. At the same time, the classification
accuracy over the original binaries was not affected by
normalization. As our experiments in \secref{sec:results}
have shown, generating functionally equivalent variants of
binaries via random transformations results in correct
classifications almost all of the time. Normalization of
binaries deterministically led to specific variants that
were correctly classified with high likelihood.

\subsection{Masking Random Instructions}
While normalization was useful for defending against
\IPR{}-based attacks, it cannot mitigate the more pernicious
\Disp{}-based attacks that are augmented with opaque or
evasive predicates. Moreover, normalization has the general
limitations that attackers could use transformations that
the normalization algorithm is not aware of or could
obfuscate code to inhibit normalization. Therefore, we
explored additional defensive measures. In particular,
motivated by the fact that randomizing binaries without the
guidance of an optimization process is unlikely to lead to
misclassification, we explored whether masking instructions
at random can mitigate attacks while maintaining high
performance on the original binaries. The defense works by
selecting a random subset of the bytes that pertain to
instructions and masking them with zeros (a commonly used
value to pad sections in binaries). While the masking is
likely to result in an ill-formed binary that is unlikely to
execute properly (if at all), the masking only occurs before
classification, which does not require a functional binary.
Depending on the classification result, one can decide
whether or not to execute the unmasked binary.

We tested the defense on binaries generated via the
\IPR{}+\Disp{}-5 white-box attack on \kaggle{} and found
that it was effective at mitigating attacks. For example,
when masking 25\% of the bytes pertaining to instructions,
the success rates of the attack decreased from 83\%--100\%
for malicious and benign binaries against the three \dnn{}s
to 0\%--20\%, while the accuracy on the original samples was
only slightly affected (e.g., it became 94\% for \ember{}).
Masking less than 25\% of the instructions' bytes was not as
effective at mitigating attacks, while masking more than
25\% led to a significant decrease in accuracy on the
original samples.

\subsection{Detecting Adversarial Examples}
To prevent binaries transformed with our attacks (i.e.,
adversarial examples) from fooling malware detection,
defenders may attempt to deploy methods to detect them. In
cases of positive detections of adversarial examples,
defenders may immediately classify them as malicious
(regardless of whether they were originally malicious or
benign). For example, because \Disp{}-based attacks increase
binaries' sizes and introduce additional \texttt{jmp}
instructions, defenders may train statistical \ml{} models
that use features such as binaries' sizes and the ratio
between \texttt{jmp} instructions and other instructions to
detect adversarial examples. While training relatively
accurate detection models may be feasible, we expect this
task to be difficult, as the attacks increase binaries'
sizes only slightly (1\%--5\%), and do not introduce many
\texttt{jmp} instructions (7\% median increase for binaries
transformed via \Disp{}-5). Furthermore, approaches for
detecting adversarial examples are likely to be susceptible
to evasion attacks (e.g., by introducing instructions after
opaque predicates to decrease the ratio between \texttt{jmp}
instructions and others). Last, another risk that defenders
should take into account is that the defense should be able
to precisely distinguish between adversarial examples and
non-adversarial benign binaries that are transformed by
similar methods to mitigate code-reuse
attacks~\cite{Koo16Rand, Pappas12Rand}.

\subsection{Takeaways}
While masking a subset of the bytes that pertain to
instructions led to better performance on adversarial
examples, it was still unable to prevent all evasion
attempts. Although the defense may raise the bar for
attackers, and make attacks even more difficult if combined
with a method to detect adversarial examples, these defenses
do not provide formal guarantees and so attackers may be
able to adapt to undermine them. For example, attackers may
build on techniques for optimization over expectations to
generate binaries that would mislead the \dnn{}s even when
masking a large number of instructions, in a similar manner
to how attackers can evade image-classification \dnn{}s
under varying lighting conditions and camera
angles~\cite{Athalye18Robust, Evtimov17Signs, Sharif16AdvML,
Sharif19AGNs}. In fact, prior work has already demonstrated
how defenses without formal guarantees are often vulnerable
to adaptive, more sophisticated,
attacks~\cite{Athalye18Attack}. Thus, since there is no
clear defense to prevent attacks against the \dnn{}s that we
studied in this work, or even general methods to prevent
attackers from fooling \ml{} models via arbitrary
perturbations, we advocate for augmenting malware-detection
systems with methods that are not based on \ml{} (e.g., ones
using templates to reason about the semantics of
programs~\cite{Christo05Semantics}), and against the use of
\ml{}-only detection methods, as has become recently
popular~\cite{Cylance}.

\section{In-Place Normalization}
\label{app:iprnorm}

\begin{figure}[!th]
  \centering
  \begin{subfigure}[t]{0.49\columnwidth}
    \includegraphics[width=\textwidth]{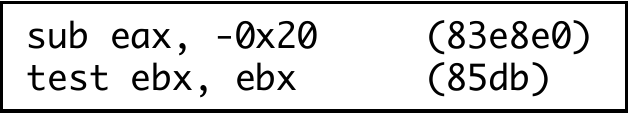}
    \caption{\label{fig:norm_original}}
  \end{subfigure}
  \begin{subfigure}[t]{0.49\columnwidth}
    \includegraphics[width=\textwidth]{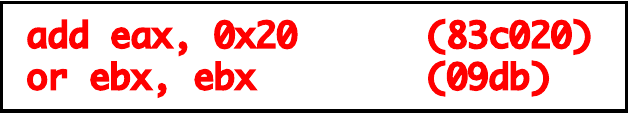}
    \caption{\label{fig:norm_equiv}}
  \end{subfigure}
  \caption{\label{fig:norm_example}An example of normalizing code via
    $\equivnt$. The original code (a) is transformed via $\equivnt$ (b) to
    decrease the lexicographic order.}
\end{figure}

In this section, we present a normalization process to map
binaries to a standard form and undo the effect of the
\IPR{} transformations on classification. Specifically, the
normalization process maps binaries to the functionally
equivalent variant with the lowest lexicographic
presentation that is achievable via the \IPR{}
transformation types. For each transformation type, we
devise an operation that would decrease a binary's
lexicographic representation when applied: \emph{1)}
instructions would be replaced with equivalent ones only if
the new instructions are lexicographically lower
($\equivnt$); \emph{2)} registers in functions would be
reassigned only if the byte representation of the first
impacted instruction would decrease ($\swap$); \emph{3)}
instructions would be reordered such that each time we would
extract the instruction from the dependence graph with the
lowest byte representation that does not depend on any of
the remaining instructions in the graph ($\reorder$); and
\emph{4)} \texttt{push} and \texttt{pop} instructions that
save register values across function calls would be
reordered to decrease the lexicographic representation while
maintaining the last-in-first-out order ($\preserv$).
\figref{fig:norm_example} depicts an example of replacing
one instruction with an equivalent one via $\equivnt$ to
decrease the lexicographic order of code.

\begin{figure}[!th]
  \centering
  
  \begin{subfigure}[t]{0.49\columnwidth}
    \includegraphics[width=\textwidth]{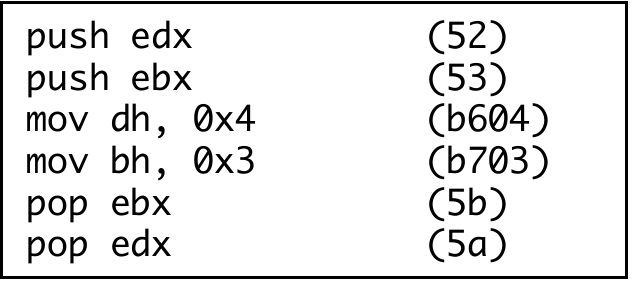}
    \caption{\label{fig:norm_is_hard:a}}
  \end{subfigure}
  \begin{subfigure}[t]{0.49\columnwidth}
    \includegraphics[width=\textwidth]{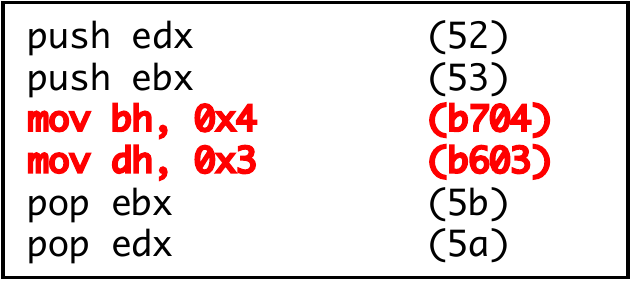}
    \caption{\label{fig:norm_is_hard:b}}
  \end{subfigure}
  
  \begin{subfigure}[t]{0.49\columnwidth}
    \includegraphics[width=\textwidth]{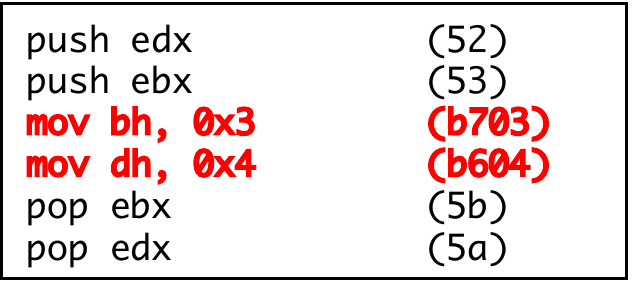}
    \caption{\label{fig:norm_is_hard:c}}
  \end{subfigure}
  \begin{subfigure}[t]{0.49\columnwidth}
    \includegraphics[width=\textwidth]{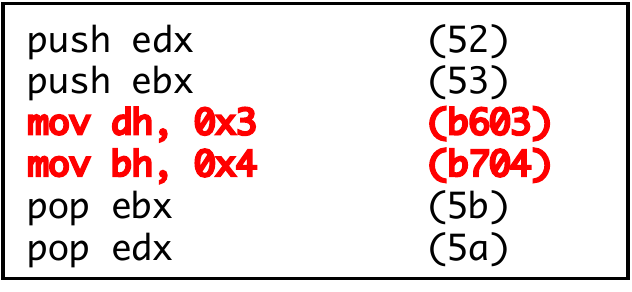}
    \caption{\label{fig:norm_is_hard:d}}
  \end{subfigure}
 
  \caption{\label{fig:norm_is_hard}The normalization process can get stuck in a
    local minima. The lexicographic order of the original code (a)
    increases when reassigning registers (b) or reordering instrcutions
    (c). However, composing the two transformation (d) decreases the
    lexicographic order.}
\end{figure}

Unfortunately, as shown in \figref{fig:norm_is_hard}, when
the different types of transformation types are composed,
applying individual normalization operations does not
necessarily lead to the binary's variant with the minimal
lexicographic representation, as the procedure may be stuck
in a local minima. To this end, we propose a stochastic
algorithm that is guaranteed to converge to binaries'
normalized variants if executed for a sufficiently large
number of iterations. 

The algorithm receives a binary $x$ and the number of
iterations $\mathit{niters}$ as inputs. It begins by drawing
a random variant of $x$, by applying all the transformation
types to each function at random. The algorithm then
proceeds to apply each of the individual normalization
operations to decrease the lexicographic representation of
the binary while self-supervising the normalization process.
Specifically, the algorithm keeps track of the last
iteration an operation decreased the binary's
representation.  
If none of the four operations affect any of the functions,
we deduce that the normalization process is stuck in a
(global or local) minima, and a random binary is drawn again
by randomizing all functions and the normalization process
restarts.

When $\mathit{niters}\rightarrow\infty$ (i.e., the number of
iterations is large enough), This algorithm would eventually
converge to a global minima. Namely, it would find the
variant of $x$ with the minimal lexicographic
representation. In fact, we are guaranteed to find
$\normFunc(x)$ even if we simply apply the transformation
types at random $x$ for $\mathit{niters}\rightarrow\infty$
iterations. When testing the algorithm with two binaries of
moderate size, we found that $\mathit{niters}$=2,000 was
sufficient to converge for the same respective variants
after every run. These variants are likely to be the global
minimas. However, executing the algorithm for 2,000
iterations is computationally expensive, and impractical
within the context of a widely deployed malware-detection
system. Hence, for the purpose of our experiments, we set
$\mathit{niters}$=10, which we found to be sufficient to
successfully mitigate the majority of attacks.

\end{document}